\documentclass[journal]{IEEEtran}
\ifCLASSINFOpdf
\else
   \usepackage{graphicx}
   \graphicspath{{../eps/}}
   \DeclareGraphicsExtensions{.eps}
\fi

\usepackage{amsmath}
\usepackage{amsfonts,amssymb}
\usepackage{amssymb}
\usepackage{wrapfig}
\usepackage{psfrag}
\usepackage{epstopdf}
\usepackage{cite}
\usepackage{graphicx}
\usepackage{subfigure}
\usepackage{threeparttable}
\usepackage{cases}
\usepackage{subeqnarray}
\usepackage{color}
\usepackage{underscore}
\usepackage{verbatim}
\usepackage{bm}
\usepackage{stfloats}
\newtheorem{theorem}{Theorem}
\newtheorem{lemma}{Lemma}
\usepackage{algorithm}
\usepackage{algorithmic}

\begin{document}

\title{Communicating with Extremely Large-Scale Array/Surface: Unified Modelling and \\ Performance Analysis}
%\title{}
%
%
%%
%%
%% author names and IEEE memberships
%% note positions of commas and nonbreaking spaces ( ~ ) LaTeX will not break
%% a structure at a ~ so this keeps an author's name from being broken across
%% two lines.
%% use \thanks{} to gain access to the first footnote area
%% a separate \thanks must be used for each paragraph as LaTeX2e's \thanks
%% was not built to handle multiple paragraphs
%%
%
\author{%Haiquan~Lu, Yong~Zeng, Shi~Jin, and Rui~Zhang
        Haiquan~Lu
        and
        Yong~Zeng,~\IEEEmembership{Member,~IEEE}
        % <-this % stops a space
\thanks{This work was supported by the National Key R\&D Program of China with Grant number 2019YFB1803400. Part of this work will be presented at the IEEE ICC 2021, Montreal, Canada, 14-23 June 2021~\cite{lu2020how}.}
\thanks{H. Lu and Y. Zeng are with the National Mobile Communications Research Laboratory, Southeast University, Nanjing 210096, China. Y. Zeng is also with the Purple Mountain Laboratories, Nanjing 211111, China (e-mail: \{haiquanlu, yong_zeng\}@seu.edu.cn). (\emph{Corresponding author: Yong Zeng.})}% <-this % stops a space

}

\maketitle


\begin{abstract}
Wireless communications with extremely large-scale array (XL-array) correspond to systems whose antenna sizes are so large that conventional modelling assumptions, such as uniform plane wave (UPW) impingement, are longer valid. This paper studies the mathematical modelling and performance analysis of XL-array communications. By deviating from the conventional modelling approach that treats the array elements as sizeless points, we explicitly model their physical area/aperture, which enables a unified modelling for the classical discrete antenna arrays and the emerging continuous surfaces. As such, a generic array/surface model that accurately takes into account the variations of signal phase, amplitude and projected aperture across array elements is proposed. Based on the proposed model, a closed-form expression of the resulting signal-to-noise ratio (SNR) with the optimal single-user maximum ratio combining/transmission (MRC/MRT) beamforming is derived. The expression reveals that instead of scaling linearly with the antenna number $M$ as in conventional UPW modelling, the SNR with the more generic model increases with $M$ with diminishing return, which is governed by the collective properties of the array, such as the \emph{array occupation ratio} and the physical sizes of the array along each dimension, while irrespective of the properties of the individual array element. In addition, we have derived an alternative insightful expression for the optimal SNR in terms of the \emph{vertical} and \emph{horizontal angular spans}, which are fully determined by the geometric angles formed by the array/surface and user location. Furthermore, we also show that our derived results include the far-field UPW modelling as a special case. One important finding during the study of far-field approximation is the necessity to introduce a new distance criterion to complement the classical Rayleigh distance, termed \emph{uniform-power distance (UPD)}, which concerns the signal amplitude/power variations across array elements, instead of phase variations as for Rayleigh distance. Extensive numerical results are provided to demonstrate the necessity of proper modelling for XL-array communications by comparing the proposed model with various benchmark models.
\end{abstract}


\IEEEpeerreviewmaketitle
\section{Introduction}
 While the commercial deployment of the fifth-generation (5G) mobile communication networks is proceeding apace, researchers worldwide have already started the investigation of beyond 5G (B5G) or sixth-generation (6G) communication networks~\cite{Zhang2020MultipleAT,Latvaaho2019Keydrivers,saad2019vision,you2021towards}. To that end, several promising transmission technologies have attracted fast-growing interest recently, such as extremely large-scale multiple-input multiple-output (XL-MIMO) communication~\cite{bjornson2019massive,marinello2020antenna}, Terahertz communication~\cite{WangJun2021General,akyildiz2016realizing}, intelligent reflecting surface (IRS) or reconfigurable intelligent surface (RIS)-assisted communications~\cite{wu2020intelligentTutorial,bjornson2020power,di2020smart,lu2021aerial,Tang2019WirelessCW}. In particular, by further increasing the antenna size/number drastically to another order-of-magnitude beyond current massive MIMO systems (typically 64 or 128 antennas only), XL-MIMO is expected to significantly improve the spectral efficiency and spatial resolution than existing systems. Besides XL-MIMO, several other terminologies are also used in the literature, such as ultra-massive MIMO (UM-MIMO)~\cite{akyildiz2016realizing}, extra-large scale massive MIMO~\cite{marinello2020antenna,decarvalho2020nonstationarities}, and extremely large aperture massive MIMO (xMaMIMO)~\cite{amiri2018extremely}. For convenience, we use the term \emph{extremely large-scale array (XL-array)} communications throughout the paper. In the extreme case when an infinite number of antenna elements are packed into a finite two-dimensional (2D) surface, XL-array converges to the emerging continuous electromagnetic (EM) surface, also known as large intelligent surfaces (LISs)~\cite{hu2018beyond,hu2018beyondpositioning,dardari2020communicating,torres2020near} or holographic MIMO surface~\cite{huang2020holographic,pizzo2020spatially}.

 Compared to the existing MIMO or massive MIMO systems, several new channel characteristics emerge when moving towards the XL-array regime. In particular, the XL-array deployed at the base station (BS), together with the continuous reduction of cell size, renders the users/scatterers less likely to be located in the far-field region~\cite{lu2020how}, where conventional uniform plane wave (UPW) assumption is usually made for ease of channel modelling and performance analysis. Note that the typical way for separating far-field versus radiative near-field regions is based on the classical Rayleigh distance ${r_{{\rm{Rayl}}}} = \frac{{2{D^2}}}{\lambda }$~\cite{selvan2017fraunhofer,yang2021communication,balanis2012advanced,balanis2016antenna}, where $D$ and $\lambda$ denote the physical dimension of the antenna array and signal wavelength, respectively. Specifically, $r_{\rm{Rayl}}$ corresponds to the minimum link distance so that if the array is used for reception, the maximum phase error of the received signals across the array elements is no greater than $\frac{\pi }{{\rm{8}}}$ by assuming \emph{normal incidence}~\cite{balanis2012advanced,balanis2016antenna}. In the far-field region with the link distance $r \geq r_{\rm{Rayl}}$, the signals can be well approximated as UPW, for which all array elements share identical signal amplitude and angle of arrival/departure (AoA/AoD) for the same channel path. However, for any given wavelength $\lambda$, as the antenna size $D$ increases, $r_{\rm Rayl}$ increases quadratically with $D$, so that the users/scatterers are more likely to be located within the Rayleigh distance. As a concrete example, for an antenna array of dimension $D=4$ meters -- a size which is not impossible for future conformal arrays (say deployed on facades of building structures), ${r_{{\rm{Rayl}}}} = 373.3$~m for signals at 3.5 GHz (frequency range 1 (FR1))~\cite{dahlman20205g}, and it is even increased to 2986.7 m when moving to 28 GHz (FR2). In such cases, near-field radiation with the more general spherical wavefront needs to be considered to more accurately model the variations of signal phase across array elements. Some preliminary efforts have been devoted towards this direction. For example, in~\cite{zhou2015spherical}, the spherical wave channel under line-of-sight (LoS) conditions was proposed. By taking into account the spherical wave propagation, a technique for realizing a high-rank channel matrix in LoS MIMO transmission was proposed in~\cite{bohagen2007design}. Apart from the planar and spherical wavefront models, an intermediate parabolic wave model was introduced in~\cite{le2019massive} to achieve a balance between model accuracy and complexity.

 Besides the need for accurate model of the signal phase relationships across array elements, XL-array communications also call for the appropriate modelling of variations of signal amplitude/power across array elements~\cite{lu2020how,bjornson2020power,hu2018beyond}, since the conventional assumption that all array elements have approximately equal distances with the terminal may no longer hold. In~\cite{lu2020how}, by taking into account the impact of amplitude variations, a closed-form expression of the received signal-to-noise ration (SNR) was derived for extremely large-scale uniform linear array (XL-ULA). It was shown that instead of scaling linearly with the antenna number $M$, the resulting SNR increases with $M$ with diminishing return, governed by a new parameter called \emph{angular span}, i.e., the angle formed by the two line segments connecting the user and both ends of the ULA. The effect of distance variations have also been considered for continuous surfaces in e.g., \cite{bjornson2020power,hu2018beyond}.

 Another new characteristic of XL-array communications is known as the spatial channel non-stationarity in complex propagation environment~\cite{decarvalho2020nonstationarities,Bian2021general}. Specifically, as the array size increases, different regions of the array may observe distinct propagation environment, e.g., different blockers and/or cluster sets, and hence exhibit different levels of received power~\cite{decarvalho2020nonstationarities}. In \cite{Payami2012Channel}, the channel measurements reported the non-stationarity over the XL-array. By characterizing such spatial non-stationarity via the concept of \emph{visibility region} of the array, a low complexity receiver architecture was proposed in \cite{amiri2018extremely}. In \cite{han2020channel}, to acquire the channel state information of the non-stationary channel in XL-array systems, the subarray-wise and scatterer-wise channel estimation methods were proposed based on visibility regions of subarrays and scatterers. Besides the method of visibility region, the non-stationarities on both the time and array dimensions were modeled as the birth-death process in \cite{wu2014non}.

 It is worth mentioning that all the aforementioned works on XL-array communications adopt the conventional modelling approach that treats the array elements as sizeless points. While such an approach is quite effective for small-to-moderate arrays, it becomes problematic for XL-arrays. Specifically, as the array size increases, signals will arrive at different array elements with quite different AoAs, i.e., the conventional assumption that all array elements share a common AoA for the same channel path no longer holds. A direct consequence is that array elements may have drastically different effective aperture to intercept the impinging wave, i.e., the projected antenna aperture that is normal to the wave propagation direction corresponding to the current element only. As will become clearer later, if such an effect is ignored, we may draw conclusions that actually violate physical laws, e.g., the received power may even exceed the transmit power when the array size grows. Note that some preliminary efforts have been devoted to taking into account the variation of projected aperture for \emph{continuous} surfaces~\cite{bjornson2020power,hu2018beyond,dardari2020communicating}, by assuming that every point on the surface is able to manipulate EM waves in real time and independently. Therefore, their obtained results cannot be applied for XL-array systems, for which EM waves can only be captured/steered by discrete array elements that are separated by certain spaces (e.g., half-wavelength spacing for classical MIMO and sub-wavelength spacing for the discrete approximation of continuous surfaces). Besides, most existing works~\cite{bjornson2020power,torres2020near} mainly consider the 2D channel modelling that only consider either the azimuth or elevation AoA/AoD, but not both as required for the more general three-dimensional (3D) channel modelling.

 To fill the above gaps, in this paper, we study the mathematical modelling and performance analysis of a generic communication system with XL-array/surface. Compared with existing relevant works such as \cite{hu2018beyond,bjornson2020power,dardari2020communicating}, our study is generic in the sense that: (i) it unifies the modelling and analysis of XL-array communications with the classical discrete array implementation and the emerging continuous surface architecture, by explicitly modelling the physical size and projected aperture of each individual element, instead of treating them as sizeless points; (ii) it pursues a generic 3D channel modelling that takes into account both zenith and azimuth AoA/AoD. By accurately modelling the variations of signal phase, amplitude and the projected aperture across array elements, a unified mathematical channel modelling that is applicable for both far-field and radiative near-field is proposed, based on which the closed-from SNR expressions for ULA- and uniform planar array (UPA)-based XL-array are derived, and some important insights in terms of the SNR scaling laws are obtained. Furthermore, a deeper study of the near- and far-field behavior of XL-array systems motivates us to propose a new distance criterion, termed \emph{uniform-power distance (UPD)}, which complements the classical Rayleigh distance for a refined near- and far-field separation. Specifically, while Rayleigh distance concerns about the phase variations across array elements, the newly proposed UPD takes into account the power or amplitude variations.
%enumerate itemize
\begin{itemize}[\IEEEsetlabelwidth{12)}]
\item Firstly, for wireless communication with extremely large-scale array/surface, a unified modelling is proposed to accurately reflect the variations of signal phase, amplitude and projected aperture across array/surface locations. Based on the proposed model, a closed-form SNR expression with the optimal single-user maximum ratio combining/transmission (MRC/MRT) beamforming is derived. The result shows that instead of scaling linearly with the antenna number $M$ as in conventional UPW modelling, the SNR with the more generic model increases with $M$ with diminishing return, which is governed by collective properties of the array, such as the so-called \emph{array occupation ratio} $\xi$, i.e., the fraction of the total array plate area that is occupied by the array elements, and its physical lengths $L_y$ and $L_z$ along each dimension, while irrespective of the individual element properties. Besides, the SNR can also be expressed in terms of the geometric angles formed by the user location and the array/surface, which we term as the \emph{horizontal} and \emph{vertical angular spans}. For the extreme case with an infinitely large planar array (not necessarily uniform) with array occupation ratio $\xi$, the resulting SNR approaches to a constant value $\frac{{\xi P}}{{2{\sigma ^2}}}$, where $P$ and $\sigma^2$ denote the transmit and noise power, respectively. This generalizes the existing result that an infinitely large continuous surface will receive half of the power transmitted by an isotropic source \cite{hu2018beyond}. Note that such intuitive results cannot be obtained if the variations of amplitude and projected aperture across array elements are not properly modelled.
\item Secondly, based on our newly derived closed-form SNR expression, we study the far-field behaviour so as to have a direct validation and comparison with the existing far-field models. It is found that the far-field approximation of the generic SNR expression shows a linear SNR increase with the antenna number $M$ or array size, which is consistent with the existing well-known results. However, our new result finds that such a linear scaling law actually depends on the user's AoA/AoD, which is usually ignored in standard MIMO models \cite{hu2018beyond} due to the ignorance of the projected aperture of array elements. Motivated by the study of near- and far-field separations, we introduce a new distance criterion, termed UPD, which concerns the signal power or amplitude variations across array elements, and it complements the classical Rayleigh distance for separating the near- and far-field regions. Furthermore, the definition of the classical Rayleigh distance is also extended to \emph{direction-dependent Rayleigh distance} such that the impact of signal directions on the phase variations across array elements is rigorously modelled.
\item Lastly, to gain further insights on our derived closed-form expressions, we focus on the special case of ULA-based architecture, for which a simpler form of the SNR expression is obtained. The result shows that the SNR for XL-ULA depends on the so-called \emph{angular span} and the \emph{angular difference}, where the cosine of the half of the angular difference reflects the impact of the projected aperture. Furthermore, we quantitatively analyze how small the ULA should be to neglect the variation of projected aperture across array elements. Extensive numerical results are provided to demonstrate the importance of taking into account both the variations of wave propagation distance and projected aperture across array elements for XL-array communications.
\end{itemize}

 The rest of this paper is organized as follows. Section \ref{sectionSystemModel} introduces the mathematical modelling for communicating with XL-array/surface. In Section \ref{SingleUserSystem}, the closed-form SNR expression for the generic UPA architecture and 3D user directions is derived, and the SNR scaling law is analyzed. In Section \ref{SectionFar-FieldApproximation}, we study the far-field behaviour of the derived expression, based on which the concepts of UPD and direction-dependent Rayleigh distance are introduced. In Section \ref{SectionUniformLinearArray}, the special case of ULA is considered and more insights are given. Numerical results are presented in Section \ref{sectionNumericalResults}. Finally, we conclude the paper in Section \ref{sectionConclusion}.

% >>>>>>>>>>>>>SECTIONS II -  here >>>>>>>>>>>>
\section{System Model}\label{sectionSystemModel}
 \begin{figure}[!t]
  \centering
  \centerline{\includegraphics[width=3.3in,height=2.75in]{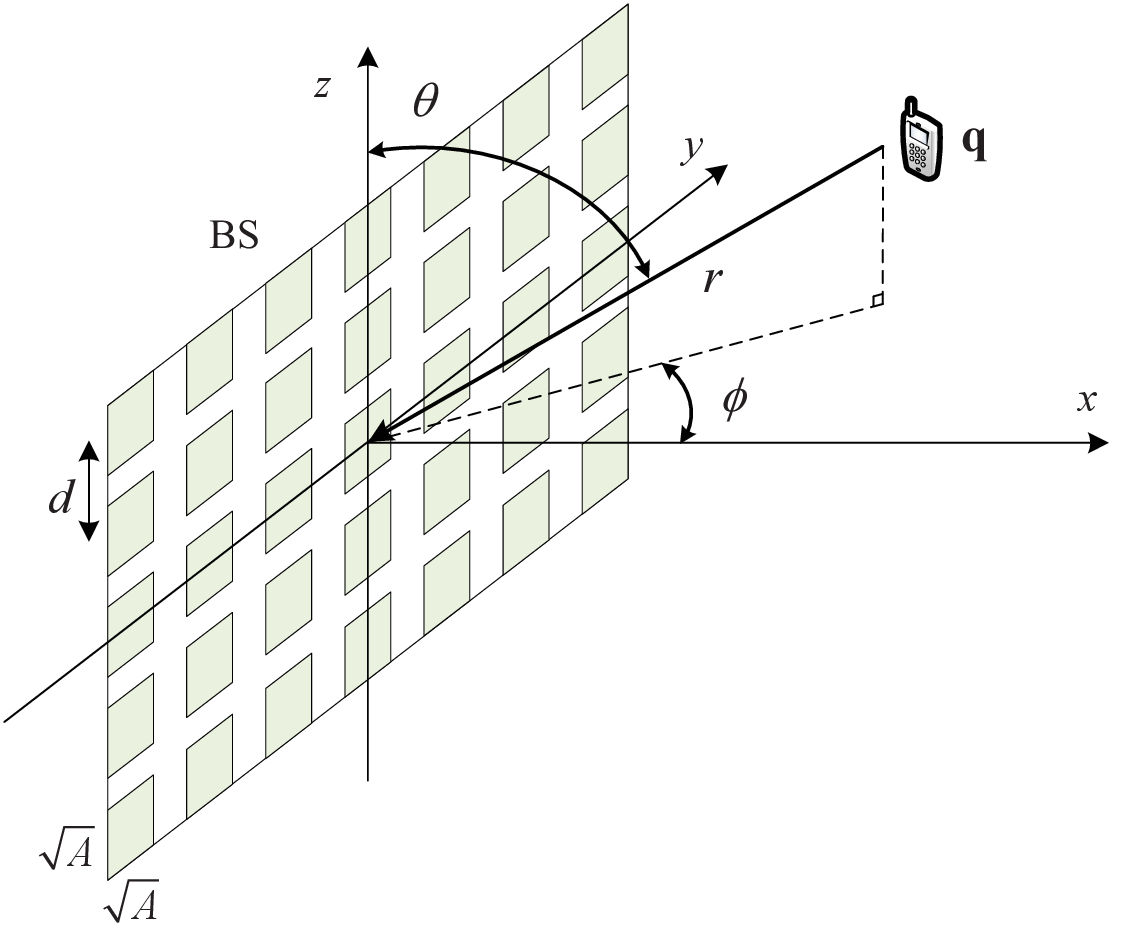}}
  \caption{Wireless communication with XL-array/surface.}
  \label{systemModel}
  \vspace{-0.5cm}
  \end{figure}
As shown in Fig.~\ref{systemModel}, we consider a wireless communication system with XL-array/surface deployed at the BS. For ease of exposition, we assume that the user has one antenna, and the array at the BS is of UPA architecture with $M \gg 1$ elements. Without loss of generality, we assume that the UPA is placed on the $y$-$z$ plane and centered at the origin, and $M = {M_y}{M_z}$, where $M_y$ and $M_z$ denote the number of antenna elements along the $y$- and $z$-axis, respectively. Different from the conventional array modelling where each array element is treated as a sizeless point, here the element size is explicitly considered, which is denoted as $\sqrt A \times \sqrt A$. As will become clear later, such a modelling is necessary to take into account the projected aperture of each array element when signals are impinged from different directions, especially when the antenna array is large. Besides, this also makes it possible to unify the modelling of discrete antenna arrays and the emerging continuous holographic surface~\cite{pizzo2020spatially,huang2020holographic}, by varying the element size $A$ and the inter-element distance $d$, where $d \ge \sqrt A $. Define $\xi  \buildrel \Delta \over = \frac{A}{{{d^2}}} \le 1$ as the \emph{array occupation ratio}, which signifies the fraction of the total UPA plate area that is occupied by the array elements. In the extreme case when $\xi  = 1$, the discrete UPA becomes a continuous holographic surface. Furthermore, let $0 < {e_a} \le 1$ denote the \emph{aperture efficiency} of each antenna element. Then the \emph{effective antenna aperture} of each element is ${A_e} = {e_a}A$. For convenience, we assume $e_a=1$ in this paper. Therefore, for the hypothetical isotropic antenna element, we have $A = A_e =\frac{{{\lambda ^2}}}{{4\pi }}$.

For notational convenience, we assume that $M_y$ and $M_z$ are odd numbers. The central location of the $\left( {{m_y},{m_z}} \right)$th array element is ${{\bf{w}}_{{m_y},{m_z}}} = {\left[ {0,{m_y}d,{m_z}d} \right]^T}$, where ${m_y} = 0, \pm 1, \cdots , \pm \left( {{M_y} - 1} \right)/2$, ${m_z} = 0, \pm 1, \cdots , \pm \left( {{M_z} - 1} \right)/2$. The physical dimensions of the UPA along the $y$- and $z$-axis are ${L_y} \approx {M_y}d$ and ${L_z} \approx {M_z}d$, respectively. For the user, let $r$ denote its distance with the center of the antenna array, and $\theta  \in \left[ {0,\pi } \right]$ and $\phi  \in \left[ { - \frac{\pi }{2},\frac{\pi }{2}} \right]$ denote the zenith and azimuth angles, respectively. Then the user location can be written as ${\bf{q}} = {\left[ {r\Psi ,r\Phi ,r\Omega } \right]^T}$ with $\Psi  \buildrel \Delta \over = \sin \theta \cos \phi$, $\Phi  \buildrel \Delta \over = \sin \theta \sin \phi$, and $\Omega  \buildrel \Delta \over = \cos \theta$. Furthermore, the distance between the user and the center of the $\left( {{m_y},{m_z}} \right)$th antenna element is
\begin{equation}\label{UserAndmthAntennaDistance}
\begin{aligned}
 {r_{{m_y},{m_z}}}& = \left\| {{{\bf{w}}_{{m_y},{m_z}}} - {\bf{q}}} \right\| \\
& = r\sqrt {1 - 2{m_y}\epsilon \Phi  - 2{m_z}\epsilon \Omega  + \left( {m_y^2 + m_z^2} \right){\epsilon ^2}},
\end{aligned}
\end{equation}
where $\epsilon  \buildrel \Delta \over = \frac{d}{r}$. Note that $r= {r_{0,0}}$ and since the array element separation $d$ is typically on the order of wavelength, in practice, we have $\epsilon \ll 1$.

Let ${S_{{m_y},{m_z}}} = \left[ {{m_y}d - \frac{{\sqrt A }}{2},{m_y}d + \frac{{\sqrt A }}{2}} \right] \times \left[ {{m_z}d - \frac{{\sqrt A }}{2},} \right.\left. {{m_z}d + \frac{{\sqrt A }}{2}} \right]$ denote the surface region of the $\left( {{m_y},{m_z}} \right)$th array element. As a theoretical analysis of the fundamental performance limits and asymptotic behaviors, we assume the basic free-space LoS propagations, for which the channel power gain between the user and the $\left( {{m_y},{m_z}} \right)$th antenna element can be written as
\begin{equation}\label{antennamChannelPowerGain}
\begin{aligned}
&{{\bar g}_{{m_y},{m_z}}}\left( {{r},{\theta},{\phi}} \right) = \\
&\int_{{S_{{m_y},{m_z}}}} {\underbrace {\frac{1}{{4\pi {{\left\| {{\bf{q}} - {\bf{s}}} \right\|}^2}}}}_{{\rm{Free-space\ path\ loss}}}\underbrace {\frac{{{{\left( {{\bf{q}} - {\bf{s}}} \right)}^T}{{{\bf{\hat u}}}_x}}}{{\left\| {{\bf{q}} - {\bf{s}}} \right\|}}}_{{\rm{Projection\ to\ signal \ direction}}}d{\bf{s}}},
\end{aligned}
\end{equation}
where ${{\bf{\hat u}}}_x$ denotes the unit vector in the $x$ direction, i.e., the normal vector of the UPA. Note that different from conventional free-space path loss modelling, the model in \eqref{antennamChannelPowerGain} further takes into account the projected aperture of each array element, as reflected by the projection of the UPA normal vector ${{\bf{\hat u}}}_x$ to the wave propagation direction at each local point $\bf{s}$. Note that \eqref{antennamChannelPowerGain} is applied to the individual array element that explicitly takes into account the element size. Similar models taking into account the projected antenna aperture have also been used in~\cite{bjornson2020power,bjornson2019utility,dardari2020communicating,hu2018beyond,hu2018beyondpositioning} for communicating with continuous surfaces. An exact evaluation of \eqref{antennamChannelPowerGain} requires a two-dimensional integration over the surface of each element. However, in practice, since the size of each individual element (not the whole array) $A$ is on the wavelength scale, the variation of the wave propagation distance ${\left\| {{\bf{q}} - {\bf{s}}} \right\|}$ and arriving direction $\frac{{{\bf{s}} - {\bf{q}}}}{{\left\| {{\bf{q}} - {\bf{s}}} \right\|}}$ across different points ${\bf{s}} \in {S_{{m_y},{m_z}}}$ is negligible. Thus, we have
\begin{equation}
\begin{aligned}
&\frac{1}{{4\pi {{\left\| {{\bf{q}} - {\bf{s}}} \right\|}^2}}} \approx \frac{1}{{4\pi {{\left\| {{\bf{q}} - {{\bf{w}}_{{m_y},{m_z}}}} \right\|}^2}}},\\
&\frac{{{{\left( {{\bf{q}} - {\bf{s}}} \right)}^T}{{{\bf{\hat u}}}_x}}}{{\left\| {{\bf{q}} - {\bf{s}}} \right\|}} \approx \frac{{{{\left( {{\bf{q}} - {{\bf{w}}_{{m_y},{m_z}}}} \right)}^T}{{{\bf{\hat u}}}_x}}}{{\left\| {{\bf{q}} - {{\bf{w}}_{{m_y},{m_z}}}} \right\|}}, \ \forall {\bf{s}} \in {S_{{m_y},{m_z}}}.
\end{aligned}
\end{equation}
As such, the channel power gain in \eqref{antennamChannelPowerGain} simplifies to
\begin{equation}\label{reducedAntennamChannelPowerGain}
\begin{aligned}
&{g_{{m_y},{m_z}}}\left( {r,{\theta},{\phi}} \right)\\
& \approx \frac{1}{{4\pi {{\left\| {{\bf{q}} - {{\bf{w}}_{{m_y},{m_z}}}} \right\|}^2}}}\underbrace {A\frac{{{{\left( {{\bf{q}} - {{\bf{w}}_{{m_y},{m_z}}}} \right)}^T}{{{\bf{\hat u}}}_x}}}{{\left\| {{\bf{q}} - {{\bf{w}}_{{m_y},{m_z}}}} \right\|}}}_{{\rm{projected\ aperture}}}\\
& = \frac{{Ar\sin {\theta}\cos {\phi}}}{{4\pi {{\left\| {{\bf{q}} - {{\bf{w}}_{{m_y},{m_z}}}} \right\|}^3}}}\\
& = \frac{{\xi \epsilon^2{\Psi }}}{{{\rm{4}}\pi {{\left[ {1 - 2{m_y}{\epsilon}{\Phi } - 2{m_z}{\epsilon}{\Omega } + \left( {m_y^2 + m_z^2} \right)\epsilon ^2} \right]}^{\frac{3}{2}}}}}.
\end{aligned}
\end{equation}
Therefore, the array response vector for the user located at distance $r$ with direction $\left( {\theta ,\phi } \right)$, denoted as ${\bf{a}}\left( {r,{\theta},{\phi}} \right)\in {{\mathbb{C}}^{M \times 1}}$, is formed by the following elements
\begin{equation}\label{newModelArrayResponseVector}
{a_{{m_y},{m_z}}}\left( {r,{\theta},{\phi}} \right) = \sqrt {{g_{{m_y},{m_z}}}\left( {r,{\theta},{\phi}} \right)} {e^{ - j\frac{{2\pi }}{\lambda }{r_{{m_y},{m_z}}}}},
\end{equation}
where ${m_y} = 0, \pm 1, \cdots , \pm \left( {{M_y} - 1} \right)/2$, ${m_z} = 0, \pm 1, \cdots , \pm \left( {{M_z} - 1} \right)/2$.

Note that different from the conventional array models, the above array response vector takes into account two new factors for XL-array: (i) the variation of wave propagation distances (and hence signal amplitude/power) across array elements; (ii) the variation of projected aperture across array elements due to the different AoAs. As a direct comparison, three commonly used array models in the existing literature are discussed below and illustrated in Fig.~\ref{UPWUSWNUSW}:

  \begin{figure}[!t]
  \centering
  \centerline{\includegraphics[width=3.6in,height=1.1in]{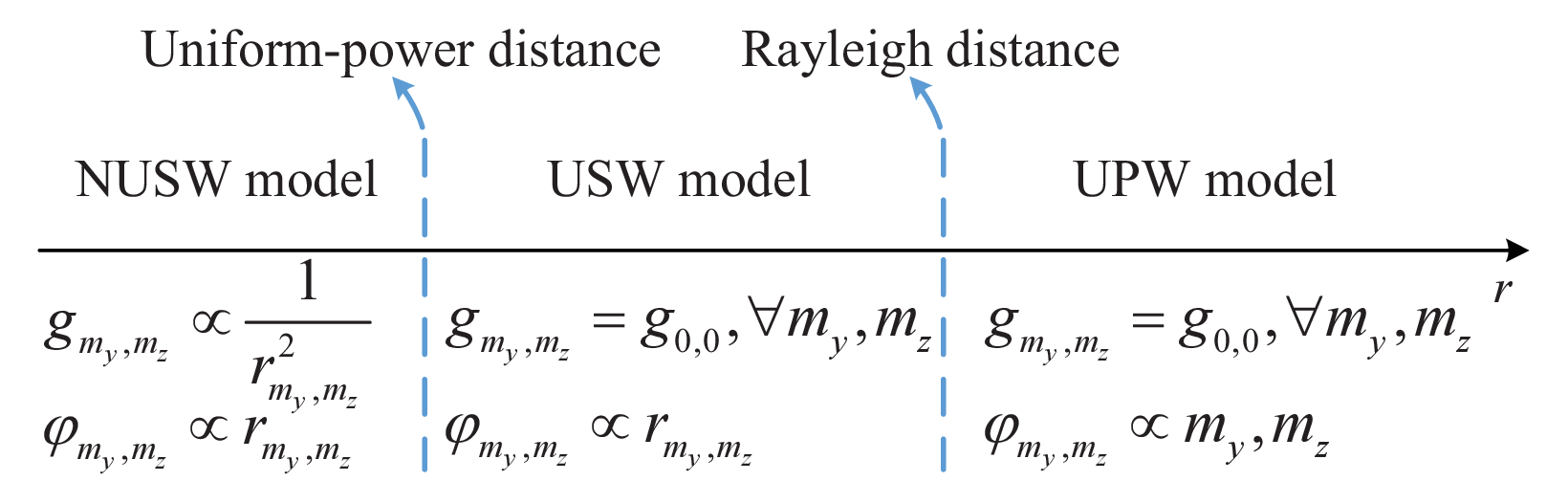}}
  \caption{An illustration of three commonly used array models, i.e., UPW, USW, and NUSW models. None of these models consider the variation of projected apertures across array elements. }
  \label{UPWUSWNUSW}
  \vspace{-0.4cm}
  \end{figure}

 1) \emph{UPW model}~\cite{bjornson2019utility,ertel1998overview}: This is the commonly used model based on the assumption that the array dimension is much smaller than the link distance, so that waves arriving at the receiver array can be well approximated as UPW. In this case, the amplitudes of the received signals by all array elements are equal, i.e., ${g_{{m_y},{m_z}}} = {g_{0,0}},\forall {m_y},{m_z}$. The variation of phase across array elements is approximated to be proportional to the separation between the element and a reference point on the array along each dimension, i.e., ${\varphi _{{m_y},{m_z}}} \propto {m_y},{m_z}$. Specifically, the $\left( {{m_y},{m_z}} \right)$th element of the array response vector for the UPW model is known as
 \begin{equation}\label{UPWArrayResponseVector}
 {a_{{m_y},{m_z}}}\left( {r,\theta ,\phi } \right) = \frac{{\sqrt {{\beta _0}} {e^{ - j\frac{{2\pi }}{\lambda }r}}}}{r}{e^{j\frac{{2\pi }}{\lambda }\left( {{m_y}d\Phi  + {m_z}d\Omega } \right)}},
 \end{equation}
 where ${\beta _0}$ denotes the channel power at the reference distance ${r_0} = 1$~m. A commonly used criterion for UPW assumption is the Rayleigh distance ${r_{{\rm{Rayl}}}} = \frac{{2{D^2}}}{\lambda }$~\cite{selvan2017fraunhofer,balanis2016antenna}.

 2) \emph{Uniform spherical wave (USW) model}~\cite{le2019massive,starer1994passive}: When the array dimension is moderately increased and/or the link distance is moderately reduced so that $r < \frac{{2{D^2}}}{\lambda }$, the ``plane wave'' assumption no longer holds, while the wave may still be ``uniform'', i.e., equal power across elements. In this case, the general spherical wave model is required to more accurately capture the variation of phase across array elements, i.e., ${\varphi _{{m_y},{m_z}}} \propto {r_{{m_y},{m_z}}}$. Thus, with the USW model, the $\left( {{m_y},{m_z}} \right)$th element of the array response vector is given by
 \begin{equation}\label{USWArrayResponseVector}
 \setlength\abovedisplayskip{0.5pt}
 \setlength\belowdisplayskip{0.5pt}
 {a_{{m_y},{m_z}}}\left( {r,\theta ,\phi } \right) = \frac{{\sqrt {{\beta _0}} }}{r}{e^{ - j\frac{{2\pi }}{\lambda }{r_{{m_y},{m_z}}}}}.
 \end{equation}

 3) \emph{Non-uniform spherical wave (NUSW) model}~\cite{zhou2015spherical,friedlander2019localization}: When the array dimension further increases and/or the link distance further reduces, even the ``uniform'' power assumption becomes invalid due to the variation of propagation distances for different array elements. In this case, both the amplitude and phase of each array element need to be modelled by its exact distance with the terminal. Specifically, the $\left( {{m_y},{m_z}} \right)$th element of the array response vector is given by
 \begin{equation}\label{NUSWArrayResponseVector}
 {a_{{m_y},{m_z}}}\left( {r,\theta ,\phi } \right) = \frac{{\sqrt {{\beta _0}} }}{{{r_{{m_y},{m_z}}}}}{e^{ - j\frac{{2\pi }}{\lambda }{r_{{m_y},{m_z}}}}}.
 \end{equation}

 It is worth mentioning that none of the above models \eqref{UPWArrayResponseVector}-\eqref{NUSWArrayResponseVector} takes into account the variation of projected aperture across array elements. As will become clear later, for XL-array communications, such an ignorance of the projected aperture variation actually leads to results that violate physical laws, e.g., the received power may even exceed the transmit power when the array dimension grows. This thus motivates our study on the more generic model in \eqref{antennamChannelPowerGain} and \eqref{newModelArrayResponseVector}. In the following, the communication performance based on the newly proposed model is analyzed, and the comparison with the existing models mentioned above is given.

 For uplink communication{\footnote[1]{Our results are directly applicable to downlink communication.}}, the received signal at the BS can be expressed as
\begin{equation}\label{BSRaeceivedSignalforUserWithoutMRC}
{\bf{y}} = {\bf{a}}\left( {r,\theta ,\phi } \right)\sqrt P s + {\bf{n}},
\end{equation}
where $P$ and $s$ are the transmit power and information-bearing signal of the user, respectively; ${\bf{n}} \sim {\cal C}{\cal N}\left( {0,{\sigma ^2}{{\bf{I}}_M}} \right)$ is the additive white Gaussian noise (AWGN). With linear receive beamforming ${\bf{v}} \in {{\mathbb{C}}^{M \times 1}}$ applied, where $\left\| {\bf{v}} \right\| = 1$, the resulting received SNR is then given by
\begin{equation}\label{ReceivedSNRForUser}
\gamma  = \frac{{P{{\left| {{{\bf{v}}^H}{\bf{a}}\left( {r,\theta ,\phi } \right)} \right|}^2}}}{{{\sigma ^2}}}.
\end{equation}
It is known that for single-user communication, the MRC beamformer is optimal, i.e., ${{\bf{v}}^*} = \frac{{{\bf{a}}\left( {r,\theta ,\phi } \right)}}{{\left\| {{\bf{a}}\left( {r,\theta ,\phi } \right)} \right\|}}$. It then follows from \eqref{reducedAntennamChannelPowerGain}, \eqref{newModelArrayResponseVector}, and \eqref{ReceivedSNRForUser} that the resulting SNR can be written as \eqref{singleUserSNRSummation}, shown at the top of the next page, where $\bar P \buildrel \Delta \over = \frac{P}{{{\sigma ^2}}}$ is the transmit SNR. Note that for the extreme case of $\theta  = 0$, $\pi$ or $\phi  =  \pm \frac{\pi }{2}$, we have $\Psi  = 0$, and it immediately follows from \eqref{singleUserSNRSummation} that $\gamma=0$. This is expected since in such extreme cases, the projected apertures of all array elements are zero. In the rest of this paper, we consider the non-trivial case where $\theta  \ne 0$, $\pi$ and $\phi  \neq  \pm \frac{\pi }{2}$.
\newcounter{mytempeqncnt1}
\begin{figure*}
\normalsize
\setcounter{mytempeqncnt1}{\value{equation}}
\begin{align} \label{singleUserSNRSummation}
\gamma  = \frac{{P{{\left\| {{\bf{a}}\left( {r,\theta ,\phi } \right)} \right\|}^2}}}{{{\sigma ^2}}} = \frac{{\bar PA\Psi }}{{4\pi {r^2}}}\sum\limits_{{m_z} =  - \frac{{{M_z} - 1}}{2}}^{\frac{{{M_z} - 1}}{2}} {\sum\limits_{{m_y} =  - \frac{{{M_y} - 1}}{2}}^{\frac{{{M_y} - 1}}{2}} {\frac{1}{{{{\left[ {1 - 2{m_y}\epsilon \Phi  - 2{m_z}\epsilon \Omega  + \left( {m_y^2 + m_z^2} \right){\epsilon ^2}} \right]}^{\frac{3}{2}}}}}} }.
\end{align}
\hrulefill
\vspace*{4pt}
\end{figure*}

% >>>>>>>>>>>>>SECTIONS III -  here >>>>>>>>>>>>
\section{Closed-Form Expression and Performance Analysis}\label{SingleUserSystem}
In this section, we derive the closed-form SNR expression of \eqref{singleUserSNRSummation} and study its scaling law.
\begin{theorem} \label{singleUserSNRapproximationTheorem}
For single-user XL-array communication with the optimal MRC/MRT beamforming, under the mild condition $r\gg d$, the resulting SNR in \eqref{singleUserSNRSummation} can be expressed in closed-form as
\begin{equation}\label{singleUserapproxiamtionSNR}
\begin{aligned}
\gamma  =& \frac{{\xi \bar P}}{{4\pi }}\left[ {U\left( {\frac{{{L_y}}}{{2r}} - \Phi ,\frac{{{L_z}}}{{2r}} - \Omega } \right)} \right. + U\left( {\frac{{{L_y}}}{{2r}} + \Phi ,\frac{{{L_z}}}{{2r}} - \Omega } \right)\\
&\left. { + U\left( {\frac{{{L_y}}}{{2r}} - \Phi ,\frac{{{L_z}}}{{2r}} + \Omega } \right) + U\left( {\frac{{{L_y}}}{{2r}} + \Phi ,\frac{{{L_z}}}{{2r}} + \Omega } \right)} \right],
\end{aligned}
\end{equation}
where $U\left( {x,y} \right){\rm{ }} \buildrel \Delta \over = \arctan \left( {\frac{{xy}}{{\Psi \sqrt {{{\Psi }^2} + {x^2} + {y^2}} }}} \right)$.
\end{theorem}

 \begin{IEEEproof}
 Please refer to Appendix~\ref{proofOfSingleUserSNRapproximationTheorem}.
 \end{IEEEproof}

Theorem~\ref{singleUserSNRapproximationTheorem} shows that with the new array response vector model formed by \eqref{newModelArrayResponseVector}, the resulting SNR no longer scales linearly with the antenna number $M$ as in conventional UPW modelling. Instead, it depends on the collective properties of the UPA in a sophisticated manner, such as the array occupation ratio $\xi$, and its physical dimensions $L_y$ and $L_z$, while irrespective of the individual element properties, such as the element size $A$ and the element separation $d$. This observation comes at no surprise, since as derived in Appendix~\ref{proofOfSingleUserSNRapproximationTheorem}, as long as the separation of each adjacent element $d$ is much smaller than the link distance $r$ (as usually the case in practice), the sum power contributed from all the array elements can be well approximated by the integration in~\eqref{approximateintegral}, and the impact of the property of the individual element becomes inconsequential. Furthermore, such observation implies that the closed-form expression \eqref{singleUserapproxiamtionSNR} is applicable not only for the UPA as shown in Fig.~\ref{systemModel}, but also for non-uniform planar arrays with potentially different element separations and/or different array sizes, as long as they are all much smaller than the link distance $r$ to validate the integration \eqref{approximateintegral}. In this case, the array occupation ratio $\xi$ is straightforwardly defined as the ratio of the total area occupied by the array elements to the total UPA plate area ${L_y}{L_z}$. Another useful observation of \eqref{singleUserapproxiamtionSNR} is that while the occupation ratio $\xi$ accounts for the total array aperture that can be used to capture the signal power, the terms inside the bracket accounts for the impact of variation of propagation distances and projected aperture across different points on the array.

To gain further insights for the closed-form expression \eqref{singleUserapproxiamtionSNR}, some special cases are considered in the following.

\begin{figure}
\centering
\subfigure[3D view]{
\begin{minipage}[t]{0.32\textwidth}
\centering
 \centerline{\includegraphics[width=3.2in,height=2.65in]{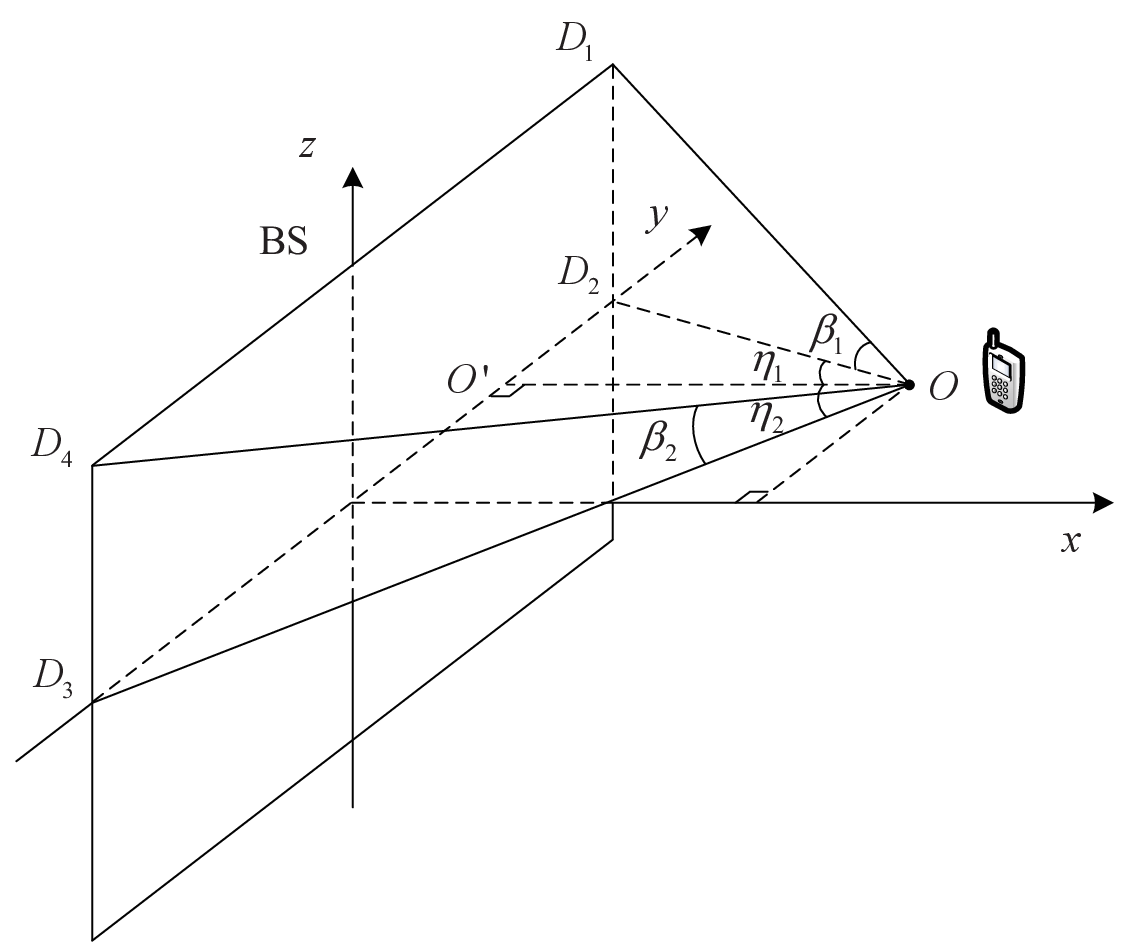}}
\end{minipage}
}
\subfigure[Top view]{
\begin{minipage}[t]{0.32\textwidth}
  \centerline{\includegraphics[width=2.35in,height=2.0in]{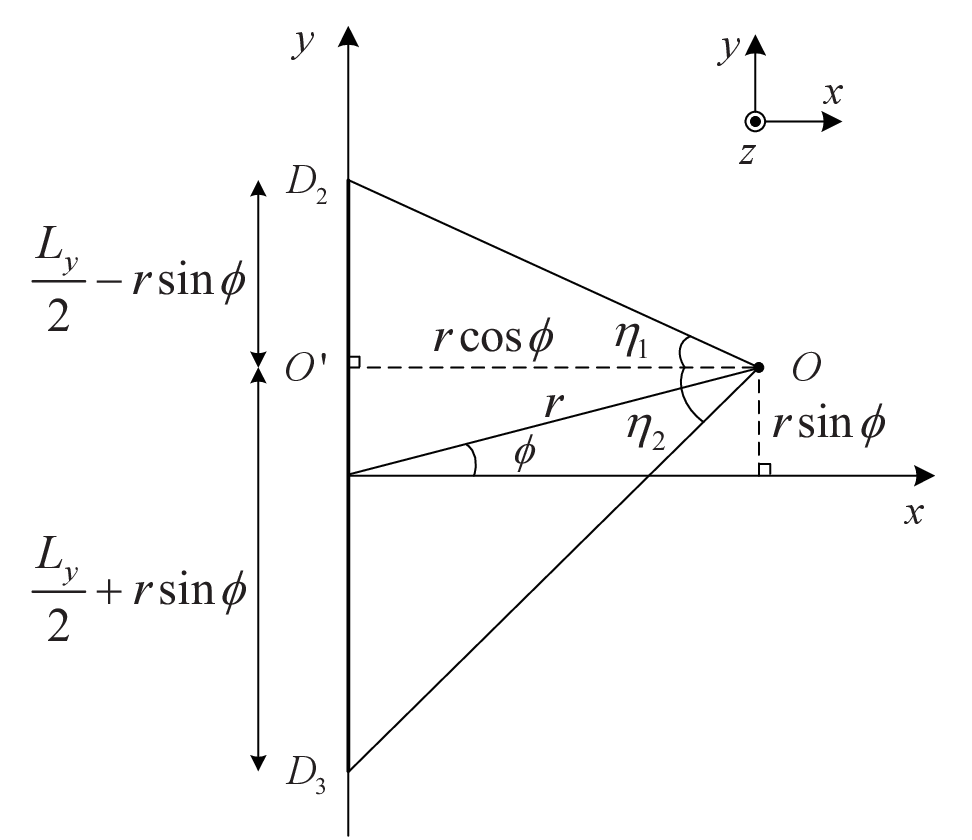}}
\end{minipage}
}
\caption{The geometrical relationships when the user is located on the $x$-$y$ plane, i.e., $\theta = \frac{\pi }{2}$.}
\label{angleIllustrationSpecialCaseThetapi2}
\end{figure}

1) $\theta  = \frac{\pi }{2}$:

When the user is located on the $x$-$y$ plane, i.e., the zenith angle $\theta  = \frac{\pi }{2}$, \eqref{singleUserapproxiamtionSNR} reduces to
\begin{equation}\label{singleUserapproxiamtionSNRtheta=pi/2}
\begin{aligned}
&\gamma  = \frac{{\xi \bar P}}{2 \pi }\left[ {\arctan \left( {\frac{{\left( {\frac{{{L_y}}}{{2r}} - \sin \phi } \right)\frac{{{L_z}}}{{2r}}}}{{\cos \phi \sqrt {{{\cos }^2}\phi  + {{\left( {\frac{{{L_y}}}{{2r}} - \sin \phi } \right)}^2} + {{\left( {\frac{{{L_z}}}{{2r}}} \right)}^2}} }}} \right)} \right.  \\
&+ \left. {\arctan \left( {\frac{{\left( {\frac{{{L_y}}}{{2r}} + \sin \phi } \right)\frac{{{L_z}}}{{2r}}}}{{\cos \phi \sqrt {{{\cos }^2}\phi  + {{\left( {\frac{{{L_y}}}{{2r}} + \sin \phi } \right)}^2} + {{\left( {\frac{{{L_z}}}{{2r}}} \right)}^2}} }}} \right)} \right].
\end{aligned}
\end{equation}
The result \eqref{singleUserapproxiamtionSNRtheta=pi/2} can be expressed as an alternative form by noting that the terms inside the two ``arctan'' functions actually depend on four geometric angles, which are fully determined by the pentahedron formed by the user location and the four corner points of the half of the UPA. Specifically, as illustrated in Fig.~\ref{angleIllustrationSpecialCaseThetapi2}, let the four corner points of the upper half of the UPA be denoted by $D_1$, $D_2$, $D_3$, and $D_4$, respectively, and the user location be denoted as $O$.  Further denote the projection of $O$ onto the $y$-axis as $O'$. Define the following four angles: ${\eta _1} = \angle {D_2}OO'$, ${\eta _2} = \angle {D_3}OO'$, ${\beta _1} = \angle {D_1}O{D_2}$, and ${\beta _2} = \angle {D_3}O{D_4}$. Then we have the following lemma.
\begin{lemma} \label{alternativeExpressionSNRtheta=pi/2lemma}
The SNR in \eqref{singleUserapproxiamtionSNR} and \eqref{singleUserapproxiamtionSNRtheta=pi/2} for $\theta  = \frac{\pi }{2}$ can be alternatively expressed as
\begin{equation}\label{alternativeExpressionSNRtheta=pi/2}
\begin{aligned}
\gamma  = \frac{{\xi \bar P}}{{2\pi }}&\left[ {{{\left( { - 1} \right)}^i}\arctan \left( {\tan {\eta _1}\sin {\beta _1}} \right)} \right.\\
&\left. { + {{\left( { - 1} \right)}^j}\arctan \left( {\tan {\eta _2}\sin {\beta _2}} \right)} \right],
\end{aligned}
\end{equation}
where
\begin{equation}\label{alternativeExpressionSNRtheta=pi/2ijvalue}
i,j = \left\{ \begin{split}
&0,1,\ \ {\rm{if}}\ r\sin \phi  <  - \frac{{{L_y}}}{2},\\
&0,0,\ \ {\rm{if}}\ - \frac{{{L_y}}}{2} \le r\sin \phi  \le \frac{{{L_y}}}{2},\\
&1,0,\ \ {\rm{if}}\ r\sin \phi  > \frac{{{L_y}}}{2}.
\end{split} \right.
\end{equation}
\end{lemma}

\begin{IEEEproof}
Please refer to Appendix \ref{proofOfalternativeExpressionSNRtheta=pi/2lemma}.
\end{IEEEproof}

Lemma~\ref{alternativeExpressionSNRtheta=pi/2lemma} shows that the optimal SNR for $\theta = \frac{\pi}{2}$ only depends on the four geometric angles formed by the user location and the UPA, which we term as the \emph{horizontal angular spans} ${\eta _1}$ and ${\eta _2}$, and the \emph{vertical angular spans} $\beta _1$ and $\beta _2$. Such an expression helps make more intuitive understanding on the resulting SNR. For example, for any given user location on the $x$-$y$ plane, as the array dimension $L_y$ increases, the horizontal angular spans $\eta_1$ and $\eta_2$ increase, which leads to higher SNR. Similar observations can be made as $L_z$ increases or the user distance $r$ decreases.

\begin{figure}[!t]
\centering
\centerline{\includegraphics[width=3.2in,height=2.65in]{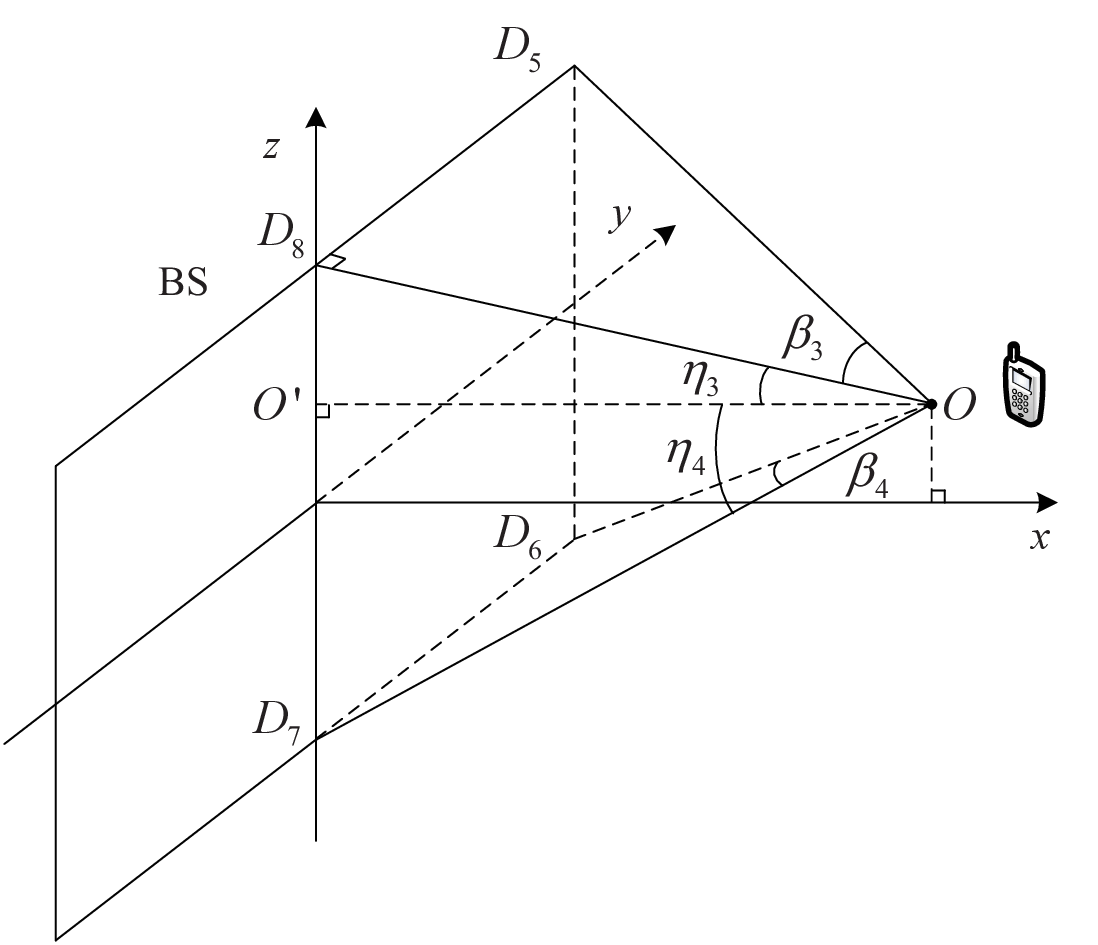}}
\caption{The geometrical relationships when the user is located on the $x$-$z$ plane, i.e., $\phi  = 0$.}
\label{angleIllustrationSpecialCasephi0}
\end{figure}

2) $\phi  = 0$:

When the user is located on the $x$-$z$ plane, i.e., the azimuth angle $\phi  = 0$, \eqref{singleUserapproxiamtionSNR} reduces to
\begin{equation}\label{singleUserapproxiamtionSNRphi=0}
\begin{aligned}
&\gamma  = \frac{{\xi \bar P}}{2 \pi }\left[ {\arctan \left( {\frac{{\frac{{{L_y}}}{{2r}}\left( {\frac{{{L_z}}}{{2r}} - \cos \theta } \right)}}{{\sin \theta \sqrt {{{\sin }^2}\theta  + {{\left( {\frac{{{L_y}}}{{2r}}} \right)}^2} + {{\left( {\frac{{{L_z}}}{{2r}} - \cos \theta } \right)}^2}} }}} \right)} \right.  \\
&+ \left. {\arctan \left( {\frac{{\frac{{{L_y}}}{{2r}}\left( {\frac{{{L_z}}}{{2r}} + \cos \theta } \right)}}{{\sin \theta \sqrt {{{\sin }^2}\theta  + {{\left( {\frac{{{L_y}}}{{2r}}} \right)}^2} + {{\left( {\frac{{{L_z}}}{{2r}} + \cos \theta } \right)}^2}} }}} \right)} \right].
\end{aligned}
\end{equation}
Similarly, as illustrated in Fig.~\ref{angleIllustrationSpecialCasephi0}, denote by $D_5$, $D_6$, $D_7$, and $D_8$ the four corner points of the right half of the UPA. By defining the following four angles ${\eta _3} = \angle {D_8}OO'$, ${\eta _4} = \angle {D_7}OO'$, ${\beta _3} = \angle {D_5}O{D_8}$, and ${\beta _4} = \angle {D_6}O{D_7}$, respectively, an alternative expression of \eqref{singleUserapproxiamtionSNRphi=0} is given in the following lemma.

\begin{lemma} \label{alternativeExpressionSNRphi=0lemma}
The SNR in \eqref{singleUserapproxiamtionSNR} and \eqref{singleUserapproxiamtionSNRphi=0} for $\phi  = 0$ can be alternatively expressed as
\begin{equation}\label{alternativeExpressionSNRphi=0}
\begin{aligned}
\gamma  = \frac{{\xi \bar P}}{{2\pi }}&\left[ {{{\left( { - 1} \right)}^i}\arctan \left( {\tan {\eta _3}\sin {\beta _3}} \right)} \right.\\
&\left. { + {{\left( { - 1} \right)}^j}\arctan \left( {\tan {\eta _4}\sin {\beta _4}} \right)} \right],
\end{aligned}
\end{equation}
where
\begin{equation}\label{alternativeExpressionSNRphi=0ijvalue}
i,j = \left\{ \begin{split}
&0,1,\ \ {\rm{if}}\ r\cos \theta  <  - \frac{{{L_z}}}{2},\\
&0,0,\ \ {\rm{if}}\ - \frac{{{L_z}}}{2} \le r\cos \theta  \le \frac{{{L_z}}}{2},\\
&1,0,\ \ {\rm{if}}\ r\cos \theta  > \frac{{{L_z}}}{2}.
\end{split} \right.
\end{equation}
\end{lemma}
\begin{IEEEproof}
The proof of Lemma~\ref{alternativeExpressionSNRphi=0lemma} is similar to that of Lemma~\ref{alternativeExpressionSNRtheta=pi/2lemma}, which is omitted for brevity.
\end{IEEEproof}

3) $\theta  = \frac{\pi }{2}$ and $\phi  = 0$:

When the user is located at the $x$-axis, i.e., $\theta  = \frac{\pi }{2}$ and $\phi  = 0$, \eqref{singleUserapproxiamtionSNR}, \eqref{singleUserapproxiamtionSNRtheta=pi/2} and \eqref{singleUserapproxiamtionSNRphi=0} reduce to
\begin{equation}\label{singleUserapproxiamtionSNRtheta=pi/2phi=0}
\gamma  = \frac{{\xi \bar P}}{\pi }\arctan \left( {\frac{{\frac{{{L_y}}}{2}\frac{{{L_z}}}{2}}}{{r\sqrt {{r^2} + {{\left( {\frac{{{L_y}}}{2}} \right)}^2} + {{\left( {\frac{{{L_z}}}{2}} \right)}^2}} }}} \right).
\end{equation}
The expression in \eqref{singleUserapproxiamtionSNRtheta=pi/2phi=0} can also be written in an alternative form by taking the special cases of \eqref{alternativeExpressionSNRtheta=pi/2} or \eqref{alternativeExpressionSNRphi=0}. Specifically, by letting ${\eta _{\rm{1}}} = {\eta _2} = \arctan \frac{{{L_y}/2}}{r}$ and ${\beta _1} = {\beta _2} = \arcsin \frac{{{L_z}/2}}{{\sqrt {{r^2} + {{\left( {{L_y}/2} \right)}^2} + {{\left( {{L_z}/2} \right)}^2}} }}$ in Fig.~\ref{angleIllustrationSpecialCaseThetapi2}(a), \eqref{singleUserapproxiamtionSNRtheta=pi/2phi=0} can also be expressed as
\begin{equation}\label{alternativeExpressionSNRtheta=pi2phi=0}
\gamma  = \frac{{\xi \bar P}}{\pi }\arctan \left( {\tan {\eta _1}\sin {\beta _1}} \right).
\end{equation}
For the given user direction $\left( {\theta ,\phi } \right) = \left( {\frac{\pi }{2},0} \right)$, as the link distance $r$ increases, both the horizontal and vertical angular spans $\eta _1$ and $\beta_1$ decrease, which leads to smaller SNR based on \eqref{alternativeExpressionSNRtheta=pi2phi=0}, as expected.

Note that by setting $A = {d^2}$ and hence $\xi  = 1$, so that the XL-array becomes a continuous surface, \eqref{singleUserapproxiamtionSNRtheta=pi/2phi=0} reduces to (19) of~\cite{hu2018beyond}. Therefore, our derived results \eqref{singleUserapproxiamtionSNRtheta=pi/2} and \eqref{singleUserapproxiamtionSNRphi=0}, and hence the more generic expression \eqref{singleUserapproxiamtionSNR}, include~\cite{hu2018beyond} as a special case.

4) ${M_y},{M_z} \to \infty $:
\begin{lemma} \label{UPAlimitValueofSNRLemma}
 For the infinitely large-scale array/surface such that ${M_y},{M_z} \to \infty $, the resulting SNR in \eqref{singleUserapproxiamtionSNR} reduces to
\begin{equation}\label{UPAlimitValueofSNR}
\mathop {\lim }\limits_{{M_y},{M_z} \to \infty } \gamma  = \frac{{\xi \bar P}}{{4\pi }} \times 4 \times \frac{\pi }{2} = \xi \frac{{\bar P}}{2}.
\end{equation}
\end{lemma}

\begin{IEEEproof}
 Lemma~\ref{UPAlimitValueofSNRLemma} can be shown by noting that the asymptotic value of the function $U\left( {x,y} \right)$ in \eqref{singleUserapproxiamtionSNR} is
 \begin{equation}\label{limitValueofUxy}
 \mathop {\lim }\limits_{x,y \to \infty } \arctan \left( {\frac{{xy}}{{\Psi \sqrt {\Psi  + {x^2} + {y^2}} }}} \right) = \frac{\pi }{2}.
 \end{equation}
\end{IEEEproof}
 This result shows that for an infinitely large array/surface with array occupation ratio $\xi$, the SNR approaches to a constant value $\frac{{\xi \bar P}}{2}$, rather than increasing unbounded as in conventional UPW models. This result makes intuitive sense since with infinitely large array, only $\frac{\xi }{2}$ of the total transmitted power will be captured. This result also generalizes the existing result that for an infinitely large \emph{continuous} surface (i.e., $\xi=1$), half of the power transmitted by an isotropic source will be captured~\cite{hu2018beyond}.

 For the hypothetical isotropic array elements with $A = \frac{{{\lambda ^2}}}{{4\pi }}$ that are separated by half-wavelength, i.e., $d = \frac{\lambda }{2}$, the array occupation ratio is $\xi {\rm{ = }}\frac{A}{{{d^2}}} = \frac{{\rm{1}}}{\pi }$. Thus we have:
\begin{lemma} \label{asymptoticBehaviorsIsotropicAntenna}
 For UPA with isotropic elements separated by half-wavelength, i.e., $A = \frac{{{\lambda ^2}}}{{4\pi }}$ and $d = \frac{\lambda }{2}$, we have
\begin{equation}\label{isotropicAntennaSNRLimit}
\mathop {\lim }\limits_{{M_y},{M_z} \to \infty } \gamma  = \frac{{\bar P}}{{2\pi }}.
\end{equation}
\end{lemma}

% >>>>>>>>>>>>>SECTIONS IV -  here >>>>>>>>>>>>
\section{Far-Field Approximation and Uniform-Power Distance}\label{SectionFar-FieldApproximation}
In this section, we study the far-field behaviour of the generic SNR expression \eqref{singleUserapproxiamtionSNR}. One important finding during the study of far-field approximation is the necessity to introduce a new distance criterion, termed UPD, to complement the classical Rayleigh distance for separating the near- and far-field propagation regions.

\begin{lemma} \label{ReduceToUPW}
When $r\sin \theta \cos \phi  \gg {L_y}$, $r\sin \theta \cos \phi  \gg {L_z}$, and $\frac{{\frac{\Phi }{\Psi }\frac{\Omega }{\Psi }}}{{\sqrt {1 + {{\left( {\frac{{{L_y}}}{{2r\Psi }} \pm \frac{\Phi }{\Psi }} \right)}^2} + {{\left( {\frac{{{L_z}}}{{2r\Psi }} \pm \frac{\Omega }{\Psi }} \right)}^2}} }} \ll 1$, the resulting SNR expression \eqref{singleUserapproxiamtionSNR} reduces to
 \begin{equation} \label{singleUserFarFieldSNR}
\gamma  \approx {\gamma _{{\rm{ff}}}} = \frac{{\bar P}}{{4\pi {r^2}}}\underbrace {MA\sin \theta \cos \phi  }_{\rm{total\ projected\ aperture}}.
 \end{equation}
 \end{lemma}
\begin{IEEEproof}
 Please refer to Appendix \ref{proofOfReduceToUPWLemma}.
 \end{IEEEproof}

Lemma~\ref{ReduceToUPW} shows that when the user is located in the far-field region, the SNR with the optimal MRC/MRT beamforming increases linearly with the antenna number $M$, which is consistent with the well-known result in the literature~\cite{bjornson2020power,ngo2013energy}. However, a new finding from \eqref{singleUserFarFieldSNR} is that such a linear scaling law also depends on the AoA/AoD via the total projected aperture $MA\sin \theta \cos \phi$, which is usually ignored in the existing literature~\cite{hu2018beyond}. Note that while the assumption of far-field with sufficiently large link distance $r$ validates the approximation that all array elements have a common AoA/AoD $\left( {\theta ,\phi } \right)$, there is no evidence that $\sin \theta \cos \phi  \approx 1$ should be satisfied since we may have highly inclined incident waves. This makes it necessary to include the term $\sin \theta \cos \phi $ as in \eqref{singleUserFarFieldSNR} accounting for the projected aperture, even for far-field approximations. As a comparison, a commonly used far-field UPW model that ignores the the impact of projected aperture is
\begin{equation}\label{farFieldModelSingleUserConventionalUPWSNR}
{\gamma _{{\rm{UPW}}}} = \bar P\frac{{M{\beta _0}}}{{{r^2}}},
\end{equation}
where $\beta_0$ is the nominal channel gain at the reference distance of $r_0 = 1$ m. For isotropic array element, ${\beta _0}  = {\left( {\frac{\lambda }{{{\rm{4}}\pi }}} \right)^{\rm{2}}}$ is usually used. By substituting $A = \frac{{{\lambda ^2}}}{{4\pi }}$ into \eqref{singleUserFarFieldSNR}, it can be found that the far-field SNR expression of the general model \eqref{singleUserFarFieldSNR} and the conventional UPW model \eqref{farFieldModelSingleUserConventionalUPWSNR} differ by $\sin \theta \cos \phi $. Therefore, the conventional far-field model \eqref{farFieldModelSingleUserConventionalUPWSNR} in general over-estimates the value in \eqref{singleUserFarFieldSNR} when the projected aperture is taken into account.

\subsection{Uniform-Power Distance}\label{subsectionCriticalDistance}
 Note that the conventional way of separating the near- and far-field regions is based on the Rayleigh (Fraunhofer) distance ${r_{\rm{Rayl}}} = \frac{{2{D^2}}}{\lambda }$~\cite{balanis2016antenna,selvan2017fraunhofer}. This is defined as the minimum distance such that the maximum phase error across array elements is no greater than $\frac{\pi}{8}$, by assuming \emph{normal incident} wave, i.e., $\left( {\theta ,\phi } \right) = \left( {\frac{\pi }{2},0} \right)$~\cite{balanis2016antenna}. In other words, the Rayleigh distance merely concerns the maximum phase difference across array elements, while irrespective of the amplitude/power difference. However, in fact, the wave propagation distance impacts the channel via both the phase and amplitude (see \eqref{newModelArrayResponseVector}). In particular, for single-user communication with the optimal MRC/MRT beamforming where the signal phases are optimally aligned, it is the amplitude variations across different elements that affect the eventual SNR, as evident from \eqref{singleUserSNRSummation}. As a consequence, the conventional way of distinguishing near- and far-field regions based on $r_{\rm Rayl}$ is insufficient. Instead, a more refined link distance partitioning that takes into account both the amplitude and phase differences across array elements is needed. To this end, we introduce a new distance criterion $r_{\rm{UPD}}$, termed UPD, beyond which the power ratio between the weakest and strongest array elements is no smaller than a certain threshold $\Gamma _{\rm{th}}$, which is illustrated in Fig.~\ref{UPWUSWNUSW}. For any given array model as a function of the link distance $r$ and direction $\left( {\theta ,\phi } \right)$, see \eqref{newModelArrayResponseVector}-\eqref{NUSWArrayResponseVector}, the power ratio between the weakest and strongest element, denoted as $\Gamma \left( {r,\theta ,\phi } \right)$, can be expressed as
 \begin{equation}\label{powerRatioDefinition}
  \setlength\abovedisplayskip{0.5pt}
 \setlength\belowdisplayskip{0.5pt}
 \Gamma \left( {r,\theta ,\phi } \right) \buildrel \Delta \over = \frac{{\mathop {\min }\limits_{{m_y},{m_z}} \ {g_{{m_y},{m_z}}}\left( {r,\theta ,\phi } \right)}}{{\mathop {\max }\limits_{{m_y},{m_z}} \ {g_{{m_y},{m_z}}}\left( {r,\theta ,\phi } \right)}}.
 \end{equation}
 It can be shown that the power ratio $\Gamma \left( {r,\theta ,\phi } \right)$ is an increasing function of the distance $r$. Therefore, for any given user direction $\left( {\theta ,\phi } \right)$, ${r_{{\rm{UPD}}}}\left( {\theta ,\phi } \right)$ is defined as the minimum link distance $r$ such that $\Gamma \left( {r,\theta ,\phi } \right)$ is no smaller than a certain threshold $\Gamma _{\rm{th}}$, i.e.,
\begin{equation}\label{rCriticalDefinition}
 \setlength\abovedisplayskip{0.5pt}
 \setlength\belowdisplayskip{0.5pt}
{r_{{\rm{UPD}}}}\left( {\theta ,\phi } \right){\rm{ }} \buildrel \Delta \over = \arg \mathop {\min }\limits_r \ \Gamma \left( {r,\theta ,\phi } \right) \ge {\Gamma _{{\rm{th}}}}.
\end{equation}
It is worth mentioning that different from the definition of Rayleigh distance~\cite{selvan2017fraunhofer,balanis2016antenna}, UPD is defined as a function of $\left( {\theta ,\phi } \right)$, and hence constitutes a surface in general. For the proposed channel power gain model, by substituting \eqref{reducedAntennamChannelPowerGain} into \eqref{powerRatioDefinition}, the power ratio follows that
\begin{equation} \label{powerRatioExpression}
\begin{aligned}
&\Gamma \left( {r,\theta ,\phi } \right) = \frac{{\mathop {\min }\limits_{{m_y},{m_z}}\  {{\left\| {{\bf{q}} - {{\bf{w}}_{{m_y},{m_z}}}} \right\|}^3}}}{{\mathop {\max }\limits_{{m_y},{m_z}}\  {{\left\| {{\bf{q}} - {{\bf{w}}_{{m_y},{m_z}}}} \right\|}^3}}}\\
 &= \frac{{{{\left[ {{r^2}{{\Psi }^2} + {{\left( {{{\left[ {r\left| {\Phi } \right| - \frac{{{L_y}}}{2}} \right]}^ + }} \right)}^2} + {{\left( {{{\left[ {r\left| {\Omega } \right| - \frac{{{L_z}}}{2}} \right]}^ + }} \right)}^2}} \right]}^{\frac{3}{2}}}}}{{{{\left[ {{r^2}{{\Psi }^2} + {{\left( {r\left| {\Phi } \right| + \frac{{{L_y}}}{2}} \right)}^2} + {{\left( {r\left| {\Omega } \right| + \frac{{{L_z}}}{2}} \right)}^2}} \right]}^{\frac{3}{2}}}}},
\end{aligned}
\end{equation}
where ${\left[ x \right]^ + } \buildrel \Delta \over = \max \left\{ {x,0} \right\}$. In particular, when $\theta  = \frac{\pi }{2}$ and $\phi = 0$, UPD is given by ${r_{{\rm{UPD}}}}\left( {\frac{\pi }{2},0} \right) = \sqrt {\frac{{\Gamma _{{\rm{th}}}^{\frac{2}{3}}}}{{1 - \Gamma _{{\rm{th}}}^{\frac{2}{3}}}}} \frac{{{L_d}}}{2}$, where ${L_d} = \sqrt {L_y^2 + L_z^2} $ denotes the diagonal dimension of the UPA. Note that the general solution ${r_{{\rm{UPD}}}}\left( {\theta ,\phi } \right)$ for \eqref{rCriticalDefinition} and \eqref{powerRatioExpression} can be obtained in closed-form, whose expression is quite sophisticated and hence omitted. Alternatively, for any given array model, the solution to \eqref{rCriticalDefinition} can be obtained numerically.

\subsection{Direction-Dependent Rayleigh Distance}\label{subsectionAngle-DependentRayleighDistance}
Motivated by the definition of UPD, in this subsection, we extend the definition of the classical Rayleigh distance to direction-dependent Rayleigh distance, so as to reflect the impact of signal direction $\left( {\theta ,\phi } \right)$ on the phase variations across array elements. Specifically, for a user with link distance $r$ and direction $\left( {\theta ,\phi } \right)$, let $\Delta \varphi \left( {r,\theta ,\phi } \right)$ denote the maximum phase error across all array elements, where the phase error of each array element is defined as the difference between its exact phase and that based on the UPW approximation as in \eqref{UPWArrayResponseVector}. Mathematically, we have
\begin{equation}\label{maximumPhaseVariation}
\Delta \varphi \left( {r,\theta ,\phi } \right) \buildrel \Delta \over = \mathop {\max }\limits_{{m_y},{m_z}} \frac{{2\pi }}{\lambda }\left[ {{r_{{m_y},{m_z}}} - \left( {r - \left( {{m_y}d\Phi  + {m_z}d\Omega } \right)} \right)} \right],
\end{equation}
To be compatible with the classical Rayleigh distance definition~\cite{balanis2016antenna}, the direction-dependent Rayleigh distance, denoted as ${r_{{\rm{ddRayl}}}}\left( {\theta ,\phi } \right)$, is defined as the minimum link distance $r$ such that $\Delta \varphi \left( {r,\theta ,\phi } \right) \le \frac{\pi }{8}$, i.e.,
\begin{equation}\label{definitionNewRayleighDistance}
{r_{{\rm{ddRayl}}}}\left( {\theta ,\phi } \right) \buildrel \Delta \over = \arg \mathop {\min }\limits_r\ \Delta \varphi \left( {r,\theta ,\phi } \right) \le \frac{\pi }{8}.
\end{equation}
It is difficult to find the closed-form solution to \eqref{definitionNewRayleighDistance} in general, but the values can be obtained numerically. Note that the definition of direction-dependent Rayleigh distance in \eqref{definitionNewRayleighDistance} generalizes the concept of Rayleigh distance to \emph{Rayleigh surface}. In particular, for the special case of $\left( {\theta ,\phi } \right) = \left( {\frac{\pi }{2},0} \right)$, we have
\begin{equation}\label{maximumPhaseVariationThetapi2Phi0}
\begin{aligned}
&\Delta \varphi \left( {r,\frac{\pi }{2},0} \right) = \mathop {\max }\limits_{{m_y},{m_z}} \frac{{2\pi }}{\lambda }\left( {r\sqrt {1 + m_y^2{\epsilon ^2} + m_z^2{\epsilon ^2}}  - r} \right)\\
& = \frac{{2\pi }}{\lambda }\left( {r\sqrt {1 + \frac{{L_d^2}}{{4{r^2}}}}  - r} \right)\mathop  \approx \limits^{\left( a \right)} \frac{{2\pi }}{\lambda }\left( {r + \frac{{L_d^2}}{{8r}} - r} \right) = \frac{{\pi L_d^2}}{{4\lambda r}},
\end{aligned}
\end{equation}
where $\left( a \right)$ follows from the first-order Taylor approximation. By substituting \eqref{maximumPhaseVariationThetapi2Phi0} into \eqref{definitionNewRayleighDistance}, we have
\begin{equation}\label{AngleDependentRayleighDistanceThetapi2Phi0}
{r_{{\rm{ddRayl}}}}\left( {\frac{\pi }{2},0} \right) = \frac{{2L_d^2}}{\lambda },
\end{equation}
which is consistent with the classical Rayleigh distance~\cite{selvan2017fraunhofer,balanis2012advanced,balanis2016antenna}.

 % >>>>>>>>>>>>>SECTIONS V -  here >>>>>>>>>>>>
\section{Uniform Linear Array}\label{SectionUniformLinearArray}
 \begin{figure}[!t]
  \centering
  \centerline{\includegraphics[width=2.9in,height=2.8in]{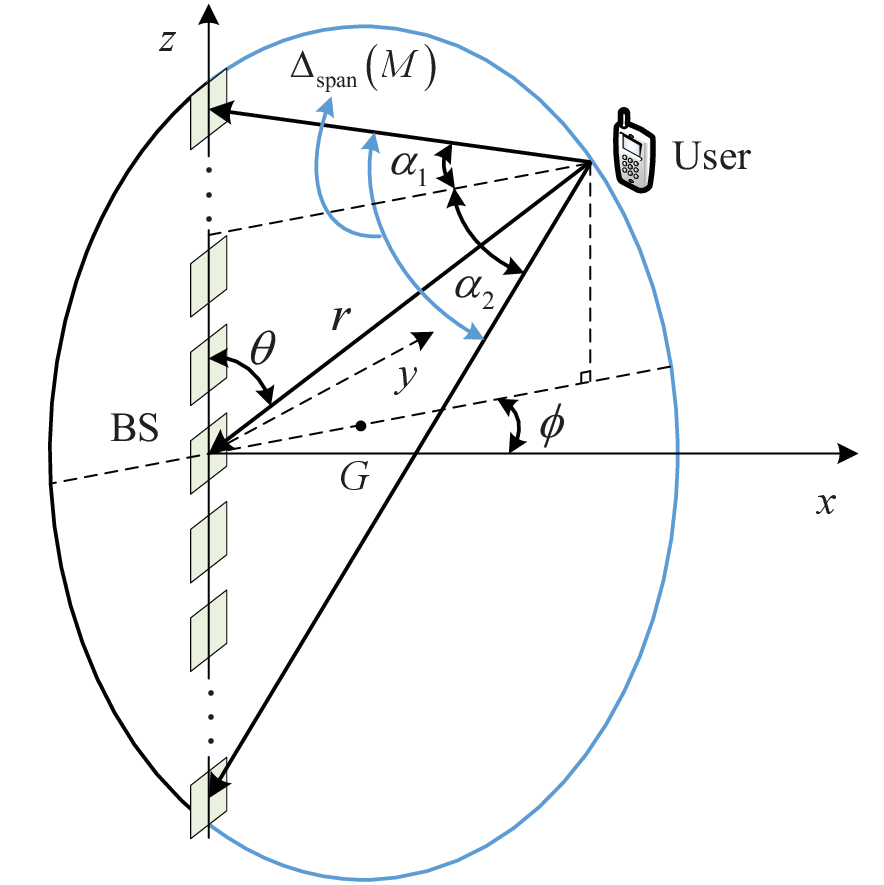}}
  \caption{The special case of ULA.}
  \label{ULAsystemModel}
  \end{figure}
 To gain further insights of the closed-form expression \eqref{singleUserapproxiamtionSNR} for the general UPA, in this section, we consider the special case of ULA, i.e., ${M_y} = 1$ and $M = {M_z}$. In this case, the closed-form expression of \eqref{singleUserapproxiamtionSNR} can be further reduced to a simpler form, which is given in the following lemma.

 \begin{lemma} \label{TransformUniformLinearArray}
 The SNR in \eqref{singleUserapproxiamtionSNR} for the special case of ULA with ${M_y} = 1$ and $M = {M_z}$ reduces to
 \begin{equation}\label{OneDimensionApproximationSNR}
 \begin{aligned}
 \gamma  &= \frac{{\bar PA\cos \phi }}{{4\pi dr\sin \theta }}\left[ {\sin \left( {{\alpha _1}\left( M \right)} \right) + \sin \left( {{\alpha _{\rm{2}}}\left( M \right)} \right)} \right]\\
 & = \frac{{\bar PA\cos \phi }}{{{\rm{2}}\pi dr\sin \theta }}\sin \left( {\frac{{{\Delta _{{\rm{span}}}}\left( M \right)}}{2}} \right)\cos \left( {\frac{{{\Delta _{{\rm{diff}}}}\left( M \right)}}{2}} \right),
 \end{aligned}
 \end{equation}
 \end{lemma}
 where ${\alpha _1}\left( M \right) \buildrel \Delta \over = \arctan \left( {\frac{{Md/2 - r\cos \theta }}{{r\sin \theta }}} \right)$, ${\alpha _{\rm{2}}}\left( M \right) \buildrel \Delta \over = \arctan \left( {\frac{{Md/2 + r\cos \theta }}{{r\sin \theta }}} \right)$, ${\Delta _{{\rm{span}}}}\left( M \right) \buildrel \Delta \over = {\alpha _1}\left( M \right) + {\alpha _{\rm{2}}}\left( M \right)$, and ${\Delta _{{\rm{diff}}}}\left( M \right) \buildrel \Delta \over = {\alpha _1}\left( M \right) - {\alpha _{\rm{2}}}\left( M \right)$.
\begin{IEEEproof}
 Please refer to Appendix \ref{proofOfTransformUniformLinearArray}.
 \end{IEEEproof}
 Lemma~\ref{TransformUniformLinearArray} shows that with the new array model that takes into account the variation of propagation distances and projected aperture across elements, the SNR for the special case of ULA depends on two new parameters, namely angular span ${\Delta _{{\rm{span}}}}\left( M \right)$ and angular difference ${\Delta _{{\rm{diff}}}}\left( M \right)$. Note that ${\Delta _{{\rm{span}}}}\left( M \right)$ is the angle formed by the two line segments connecting the user and both ends of the antenna array, as illustrated in Fig.~\ref{ULAsystemModel}. It can be shown that as $M \to \infty$, $\sin \left( {{\alpha _1}\left( M \right)} \right) + \sin \left( {{\alpha _{\rm{2}}}\left( M \right)} \right) \to 2$. This thus leads to
 \begin{equation}\label{ULAlimitValueofSNR}
 \mathop {{\rm{lim}}}\limits_{M \to \infty } \gamma  = \frac{{\bar PA\cos \phi }}{{2\pi dr\sin \theta }},
 \end{equation}
 which is a constant that depends on the user's projected distance to the ULA $r\sin \theta$ and the projected aperture $A\cos \phi $. Note that this result is different from the asymptotic limit of UPA in \eqref{UPAlimitValueofSNR}, for which $\frac{\xi }{2}$ of the transmitted power is captured. This is expected since the array aperture of the one-dimensional infinitely long ULA is much smaller than that of the two-dimensional UPA.

 In our preliminary work~\cite{lu2020how}, the SNR of ULA is derived by only taking into account the distance variations, but ignoring the projected element apertures, which leads to \cite{lu2020how}
 \begin{equation}\label{conventionalModelOneDimensionApproximationSNR}
 {\gamma _{{\rm{NUSW,ULA}}}} = \bar P\frac{{{\lambda ^2}}}{{{{\left( {4\pi } \right)}^2}dr\sin \theta }}{\Delta _{{\rm{span}}}}\left( M \right),
 \end{equation}
 which only depends on the angular span and is irrelevant to the angular difference as in \eqref{OneDimensionApproximationSNR}. To show the impact of ${\Delta _{{\rm{diff}}}}\left( M \right)$ in \eqref{OneDimensionApproximationSNR}, we consider the scenario shown in Fig.~\ref{ULAsystemModel}. Specifically, for a given user location with distance $r$ and direction $\left( {\theta ,\phi } \right)$, we form a circle that passes through the two end points of the ULA and the user location. Therefore, the chord formed by the ULA divides the circle into two parts. According to the geometric relationship, the center $G$ and the radius of the circle can be given by ${\left[ {\frac{{Md}}{2}{\rm{cot}}\left( {{\Delta _{{\rm{span}}}}\left( M \right)} \right)\cos \phi ,\frac{{Md}}{2}{\rm{cot}}\left( {{\Delta _{{\rm{span}}}}\left( M \right)} \right)\sin \phi ,0} \right]^T}$ and $\frac{{Md/2}}{{\sin \left( {{\Delta _{{\rm{span}}}}\left( M \right)} \right)}}$, respectively. Furthermore, according to circle theorems, when the user moves along the right blue arc on the circle, it will have a constant circumferential angle, i.e., the angular span ${\Delta _{{\rm{span}}}}\left( M \right)$, but different angular difference ${\Delta _{{\rm{diff}}}}\left( M \right)$.

  \begin{figure}[!t]
  \centering
  \centerline{\includegraphics[width=3.5in,height=2.625in]{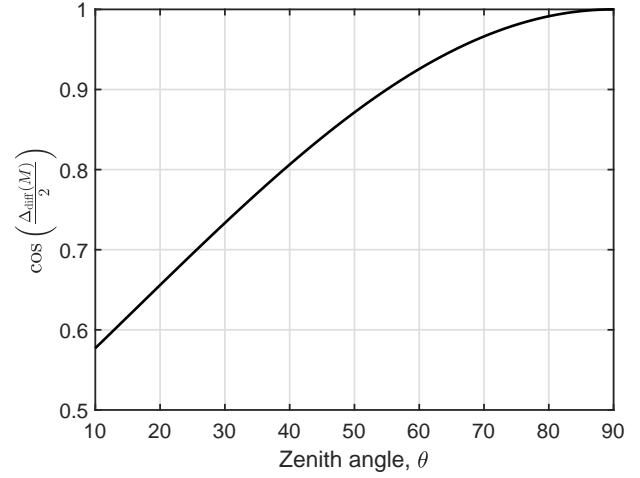}}
  \caption{The term $\cos \left( {\frac{{{\Delta _{{\rm{diff}}}}\left( M \right)}}{2}} \right)$ versus zenith angle $\theta$.}
  \label{cosineItemVersusElevationAngle}
 \end{figure}

 Under the above setup, for a fixed angular span ${\Delta _{{\rm{span}}}}\left( M \right) = \frac{\pi }{3}$, Fig.~\ref{cosineItemVersusElevationAngle} shows the term $\cos \left( {\frac{{{\Delta _{{\rm{diff}}}}\left( M \right)}}{2}} \right)$ in \eqref{OneDimensionApproximationSNR} versus the user's zenith angle $\theta$ as the user moves along the right part of the circle in Fig.~\ref{ULAsystemModel}. It is observed that as $\theta$ increases, $\cos \left( {\frac{{{\Delta _{{\rm{diff}}}}\left( M \right)}}{2}} \right)$ increases, and it is equal to one for $\theta=\frac{\pi}2$. This is expected since when $\theta$ increases from $0$ to $\frac{\pi}2$, the projected aperture of antenna elements increases. Therefore, $\cos \left( {\frac{{{\Delta _{{\rm{diff}}}}\left( M \right)}}{2}} \right)$ in \eqref{OneDimensionApproximationSNR} can be regarded as a reflection of the variation of the projected aperture impacted by the user direction.

 Furthermore, a closer look at \eqref{antennamChannelPowerGain} reveals that both the free-space path loss and projected aperture of each array element affect the channel power gain of the array elements. To analyze the effects of these two separate factors on the resulting power gain, let ${\rho _{{\rm{PL}}}}{\rm{ }} \buildrel \Delta \over = \frac{{{{\left\| {{\bf{q}} - {{\bf{w}}_{\left( {M - 1} \right){\rm{/2}}}}} \right\|}^2}}}{{{{\left\| {{\bf{q}} - {{\bf{w}}_0}} \right\|}^2}}}$, ${\rho _{{\rm{aper}}}} \buildrel \Delta \over = \frac{{{{\left( {{\bf{q}} - {{\bf{w}}_{\left( {M - 1} \right){\rm{/2}}}}} \right)}^T}{{{\bf{\hat u}}}_x}/\left\| {{\bf{q}} - {{\bf{w}}_{\left( {M - 1} \right){\rm{/2}}}}} \right\|}}{{{{\left( {{\bf{q}} - {{\bf{w}}_0}} \right)}^T}{{{\bf{\hat u}}}_x}/\left\| {{\bf{q}} - {{\bf{w}}_0}} \right\|}}$ and ${\rho} \buildrel \Delta \over = {\rho _{{\rm{PL}}}}{\rho _{{\rm{aper}}}}$ be the normalized free-space path loss, projected aperture and channel power gain of the array element for the last element of the ULA with respect to the central element, respectively.

  \begin{figure}[!t]
  \centering
  \centerline{\includegraphics[width=3.5in,height=2.625in]{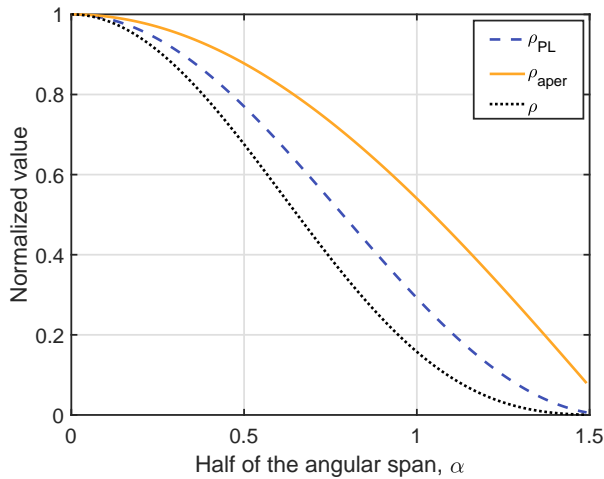}}
  \caption{Normalized free-space path loss, projected aperture and channel power gain versus half of the angular span in radians.}
  \label{pathGainProjectionVersusAngularSpan}
 \end{figure}

 Fig.~\ref{pathGainProjectionVersusAngularSpan} plots the above factors versus half of the angular span $\alpha$ when $\left( {\theta ,\phi } \right) = \left( {\frac{\pi }{2},0} \right)$, where $\alpha_1 = \alpha_2 = \alpha  = \arctan \frac{{Md/2}}{r}$. It is observed that as the angular span increases, i.e., with larger array or shorter link distance, both the normalized free-space path loss and projected aperture decrease, which lead to smaller contribution of the channel power gain by the end elements along the ULA, as compared to those by the central element. It is also observed from Fig.~\ref{pathGainProjectionVersusAngularSpan} that the impact of free-space path loss caused by distance variations is more significant than that caused by the projected aperture arising from the AoA variations.
\section{Numerical Results}\label{sectionNumericalResults}
In this section, numerical results are provided to compare the various models for XL-array communications. Unless otherwise stated, the transmit SNR is $\bar P = 90$~dB, and the antenna separation is set as $d = \frac{\lambda }{2}=0.0628$~m. The distance between the user and the center of the antenna array is $r = 25$~m, and the size of each array element is $A = \frac{{{\lambda ^2}}}{{4\pi }}$.

\subsection{Comparison of Different Array Models}
 \begin{figure}
\centering
\subfigure[$\left( {\theta ,\phi } \right) = \left( {\frac{\pi }{6},\frac{\pi }{3}} \right)$]{
\begin{minipage}[t]{0.5\textwidth}
\centering
\centerline{\includegraphics[width=3.5in,height=2.625in]{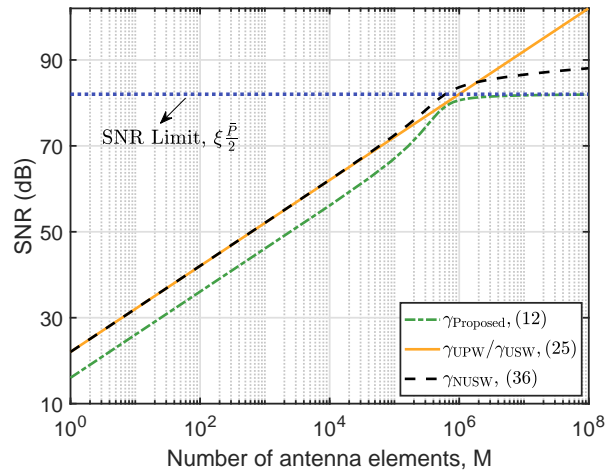}}
\end{minipage}
}
\subfigure[$\left( {\theta ,\phi } \right) = \left( {\frac{\pi }{2},\frac{\pi }{4}} \right)$]{
\begin{minipage}[t]{0.5\textwidth}
\centering
\centerline{\includegraphics[width=3.5in,height=2.625in]{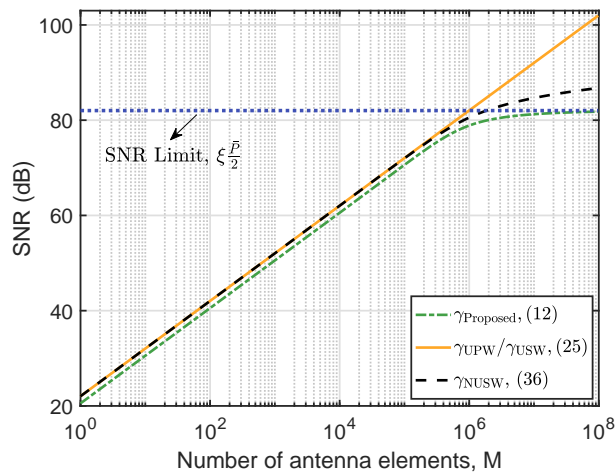}}
\end{minipage}
}
\caption{Comparison of different models versus antenna number $M$ for UPA.}
\label{summationVersusIntegralLog}
\end{figure}
 First, we compare the resulting SNR for the four array models, i,e., the proposed general model in \eqref{newModelArrayResponseVector}, the UPW model in \eqref{UPWArrayResponseVector}, the USW model in \eqref{USWArrayResponseVector}, and the NUSW model in \eqref{NUSWArrayResponseVector}. Note that with the optimal MRC/MRT beamforming where the phases are properly aligned, the resulting SNR of the UPW and USW models are identical. In addition, by substituting \eqref{NUSWArrayResponseVector} into \eqref{ReceivedSNRForUser}, the received SNR for the NUSW model can be calculated based on the following formula \cite{lu2020how}
 \begin{equation}\label{NUSWSNRForUser}
 \setlength\abovedisplayskip{0.5pt}
 \setlength\belowdisplayskip{0.5pt}
  \gamma _{\rm{NUSW}} = \bar P\sum\limits_{{m_z}} {\sum\limits_{{m_y}} {\frac{{{\beta _{\rm{0}}}{\rm{/}}{r^2}}}{{1 - 2{m_y}\epsilon  \Phi  - 2{m_z}\epsilon \Omega  + \left( {m_y^2 + m_z^2} \right){\epsilon ^2}}}} }.
 \end{equation}

 Fig.~\ref{summationVersusIntegralLog} plots the resulting SNR $\gamma_{\rm{Proposed}}$, $\gamma _{\rm{UPW}}$/$\gamma _{\rm{USW}}$, and $\gamma _{\rm{NUSW}}$ versus the number of antenna elements $M$ for a square UPA (i.e., $M_y = M_z$) with user directions $\left( {\theta ,\phi } \right) = \left( {\frac{\pi }{6},\frac{\pi }{3}} \right)$ and $\left( {\theta ,\phi } \right) = \left( {\frac{\pi }{2},\frac{\pi }{4}} \right)$, respectively. The asymptotic SNR limit \eqref{UPAlimitValueofSNR} is also shown in the figure. It is firstly observed that for moderate antenna number $M$, the SNR for all models increases linearly with $M$, which is in accordance with Lemma~\ref{ReduceToUPW} and \eqref{farFieldModelSingleUserConventionalUPWSNR}. However, the three existing models that ignore the aperture effect of array elements, i.e., $\gamma _{\rm{UPW}}$/$\gamma _{\rm{USW}}$ and $\gamma _{\rm{NUSW}}$ in general over-estimate the true value predicted by the proposed model, and the gap becomes more significant with inclined wave incidence, i.e., when $\sin \theta \cos \phi$ is small. This observation is consistent with the expressions \eqref{singleUserFarFieldSNR} and \eqref{farFieldModelSingleUserConventionalUPWSNR}. Furthermore, as $M$ increases, $\gamma_{\rm{Proposed}}$ and $\gamma _{\rm{UPW}}$/$\gamma _{\rm{USW}}$ exhibit drastically different scaling laws, i.e., approaching to a constant value versus increasing linearly and unbounded. It is also observed that the NUSW model, which takes into account the power variation across array elements, also exhibits diminishing return for extremely large $M$. However, due to the ignorance of the impact of projected apertures, the asymptotic value of $\gamma _{\rm{NUSW}}$ even exceeds $\xi \frac{{\bar P}}{2}$, which is impossible, since the maximum power that could be captured by the UPA is $\xi \frac{P}{2}$. These observations demonstrate the importance of properly modelling both the variations of signal power and projected apertures across array elements, as in our proposed model.

 \begin{figure}[!t]
  \centering
  \centerline{\includegraphics[width=3.5in,height=2.625in]{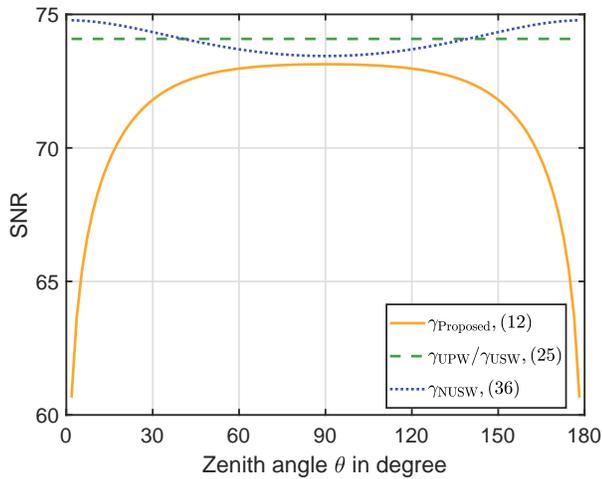}}
  \caption{Comparison of different models versus user zenith angle $\theta$, where $\phi=0$.}
  \label{SNRVersusUserZenithAngleUPA}
 \end{figure}

 As a further illustration, Fig.~\ref{SNRVersusUserZenithAngleUPA} shows the resulting SNR $\gamma_{\rm{Proposed}}$, $\gamma _{\rm{UPW}}$/$\gamma _{\rm{USW}}$, and $\gamma _{\rm{NUSW}}$ versus the zenith angle $\theta$ for a square UPA, where the azimuth angle is fixed to $\phi=0$. The number of antenna elements is $M_y = M_z = 400$. It is observed that while the resulting SNR of the UPW/USW models is a constant as $\theta$ changes, that for the NUSW and the proposed models critically depend on the user direction. Specifically, as $\theta$ increases from $0$ to $\frac{\pi}{2}$, $\gamma _{\rm{NUSW}}$ monotonically decreases, while the opposite trend is observed with the proposed model. This is expected since the variation of $\gamma _{\rm{NUSW}}$ versus $\theta$ is only due to the distance variations across array elements, while that for $\gamma_{\rm{Proposed}}$ takes into account variations of both distances and projected aperture. When $\sin \theta $ is small, the projected aperture of the array element is quite small, as can be seen from \eqref{reducedAntennamChannelPowerGain}, which is ignored in existing models and thus lead to significant performance gaps between different models.

\subsection{UPD and Direction-Dependent Rayleigh Distance}

\begin{figure}[!t]
  \centering
  \centerline{\includegraphics[width=3.5in,height=2.625in]{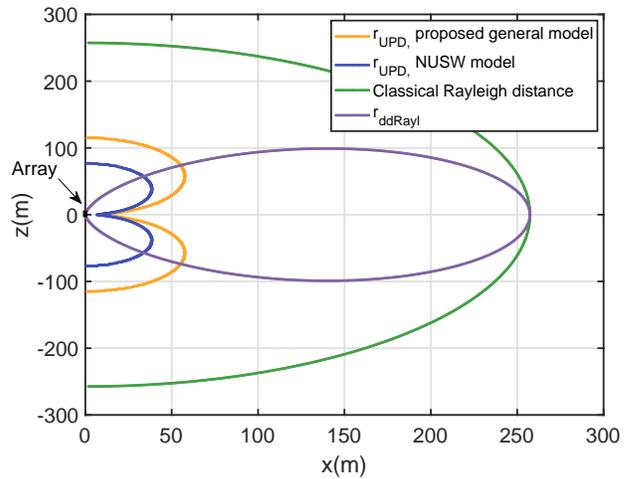}}
  \caption{Comparison of UPD, classical Rayleigh distance, and direction-dependent Rayleigh distance.}
  \label{partitionOfNearAndFarField}
 \end{figure}

 Next, the UPD and direction-dependent Rayleigh distance introduced in Section~\ref{subsectionCriticalDistance} and \ref{subsectionAngle-DependentRayleighDistance} are illustrated. For ease of exposition, an ULA-based array with $M = M_z = 64$ is considered. With the user's azimuth angle fixed to $\phi = 0$ while varying the zenith angle $\theta$, the four corresponding distance curves are plotted in Fig.~\ref{partitionOfNearAndFarField} based on definitions \eqref{rCriticalDefinition} and \eqref{definitionNewRayleighDistance}. Note that for UPD, both the proposed model \eqref{newModelArrayResponseVector} and the NUSW model \eqref{NUSWArrayResponseVector} are considered, with the power ratio threshold set to $\Gamma _{\rm{th}} = 90\%$. It is observed from Fig.~\ref{partitionOfNearAndFarField} that the four distance curves are quite different, due to the different array models and criteria used. For example, while the classical Rayleigh distance results in a semicircle, due to its definition based on normal incidence assumption, that for direction-dependent Rayleigh distance exhibits an ellipse shape. In particular, while consistent results are obtained for the special case of normal incidence (i.e., $\theta = \frac{\pi}{2}$), for general direction with $\theta  \ne \frac{\pi }{{\rm{2}}}$, the classical Rayleigh distance is in fact a conservative criterion for far-field approximation from the phase modelling perspective.
 
 Fig.~\ref{partitionOfNearAndFarField} also shows that for UPD introduced in \eqref{rCriticalDefinition}, the proposed and NUSW models lead to quite different values. For both array models, UPD achieves the minimum when $\theta  = \frac{\pi}{2}$, and for inclined direction with $\theta  \ne \frac{\pi }{{\rm{2}}}$, UPD increases significantly. This implies that if UPD is used for refined near- and far-field separations, users with inclined direction are more likely to be located within the near-field region. Thus, the conclusion made in \cite{bjornson2020power} that far-field approximation is safely applied in many practical scenarios may not apply here, since it was drawn only based on normal incidence assumption. By further comparing UPD and direction-dependent Rayleigh distances in Fig.~\ref{partitionOfNearAndFarField}, the exactly opposite trends are observed as the user direction $\theta$ varies. For direction-dependent Rayleigh distance criterion, when $\theta  = \frac{\pi}{2}$, the far-field distance is the largest while it is the smallest for UPD criterion. Such results demonstrate the importance of proper channel power model for XL-array and the necessity to introduce UPD and direction-dependent Rayleigh distance to capture the variations of the signal power and phase across array elements.

\subsection{Uniform Linear Array}
 \begin{figure}[!t]
  \centering
  \centerline{\includegraphics[width=3.5in,height=2.625in]{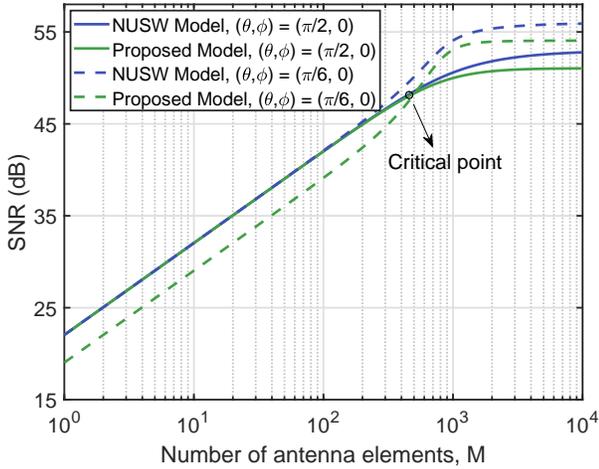}}
  \caption{SNR versus antenna number $M$ for ULA.}
  \label{SNRVersusAngularSpan}
 \end{figure}
  Last, we consider the special case of ULA in Section~\ref{SectionUniformLinearArray}. Fig.~\ref{SNRVersusAngularSpan} shows the SNR versus the number of antenna elements $M$ with the proposed and NUSW models. The critical point where the proposed model reduces to the NUSW model for normal user direction $\left( {\theta ,\phi } \right) = \left( {\frac{\pi }{2},0} \right)$ is also labelled. Specifically, the critical point is defined as the maximum antenna number such that the SNR ratio between $\gamma$ in \eqref{OneDimensionApproximationSNR} and $\gamma _{{\rm{NUSW,ULA}}}$ in \eqref{conventionalModelOneDimensionApproximationSNR} is no smaller than the threshold 95\%. It is observed that even for normal incidence with $\left( {\theta ,\phi } \right) = \left( {\frac{\pi }{2},0} \right)$, as the number of antenna elements exceeds the critical point, the two different models lead to different SNR predictions. In fact, the performance deviation is even more significant for inclined user directions, e.g., $\left( {\theta ,\phi } \right) = \left( {\frac{\pi }{6},0} \right)$, even for small number of array elements. This is due to the fact that the proposed model takes into account the projected aperture, while it is ignored by the NUSW model, as can be seen from \eqref{OneDimensionApproximationSNR} and \eqref{conventionalModelOneDimensionApproximationSNR}. As a consequence, the NUSW model in general over-estimates the value predicted by the general model.

 % >>>>>>>>>>>>>SECTIONS VII -  here >>>>>>>>>>>>
\section{Conclusion}\label{sectionConclusion}
This paper studied the mathematical modelling and performance analysis for communicating with XL-array/surface. A unified modelling approach was proposed for discrete array and continuous surface, by taking into account the variations of signal phase, power and projected aperture across array elements. With the optimal MRC/MRT beamforming, a closed-form SNR expression was derived for single-user uplink/downlink communication with 3D user directions, based on which some important insights were obtained. We further analyzed the far-field behavior of the derived SNR expression and introduced a new distance criterion termed UPD, together with the extension of the classical Rayleigh distance to direction-dependent Rayleigh distance. Extensive numerical results were provided to demonstrate the importance of proper modelling for XL-array communications.

% >>>>>>>>>>>>> appendices -  here >>>>>>>>>>>>
\begin{appendices}
\section{Proof of Theorem~\ref{singleUserSNRapproximationTheorem}}\label{proofOfSingleUserSNRapproximationTheorem}
Based on the SNR expression \eqref{singleUserSNRSummation}, we first define the function $f\left( {y,z} \right) \buildrel \Delta \over = \frac{1}{{{{\left( {1 - 2 y \Phi - 2 z \Omega + {y^2} + {z^2}} \right)}^{\frac{3}{2}}}}}$ over the rectangular area ${\mathcal A} = \left\{ {\left( {y,z} \right)\left| { - \frac{{{M_y}\epsilon}}{2} \le y \le \frac{{{M_y}\epsilon}}{2}, - \frac{{{M_z}\epsilon}}{2} \le z \le \frac{{{M_z}\epsilon}}{2}} \right.} \right\}$. $\mathcal{A}$ is then partitioned into ${M_y}{M_z}$ subrectangles, each of equal area $\epsilon^2$. Since $\epsilon \ll 1$ in practice, we have $f\left( {y,z} \right) \approx f\left( {{m_y}\epsilon ,{m_z}\epsilon } \right),\forall \left( {y,z} \right) \in \left[ {\left( {{m_y} - \frac{1}{2}} \right)\epsilon ,\left( {{m_y} + \frac{1}{2}} \right)\epsilon } \right] \times \left[ {\left( {{m_z} - \frac{1}{2}} \right)\epsilon ,\left( {{m_z} + \frac{1}{2}} \right)\epsilon } \right]$. Then based on the concept of double integral, we have
\begin{equation}\label{approximateintegral}
\sum\limits_{{m_z} =  - \frac{{{M_z} - 1}}{2}}^{\frac{{{M_z} - 1}}{2}} {\sum\limits_{{m_y} =  - \frac{{{M_y} - 1}}{2}}^{   \frac{{{M_y} - 1}}{2}} {f\left( {{m_y}\epsilon ,{m_z}\epsilon } \right){\epsilon ^2}} }  \approx \iint_{\mathcal A}{f\left( {y,z} \right)dydz},
\end{equation}
By substituting $f\left( {y,z} \right)$ into \eqref{approximateintegral} and taking the double integral, we have \eqref{Integrate1}, shown at the top of the next page, where ${\left( a \right)}$ and ${\left( b \right)}$ follow from the integral formulas 2.264.5 and 2.284 in~\cite{gradshteyn2014table}, respectively.
\newcounter{mytempeqncnt3}
\begin{figure*}%公式位置按图放置调整
\normalsize
\setcounter{mytempeqncnt3}{\value{equation}}
\begin{align} \label{Integrate1}
&\sum\limits_{{m_z} =  - \frac{{{M_z} - 1}}{2}}^{\frac{{{M_z} - 1}}{2}} {\sum\limits_{{m_y} =  - \frac{{{M_y} - 1}}{2}}^{\frac{{{M_y} - 1}}{2}} {\frac{1}{{{{\left[ {1 - 2 {m_y}\epsilon \Phi   - 2 {m_z}\epsilon \Omega + \left( {m_y^2 + m_z^2} \right){\epsilon ^2}} \right]}^{\frac{3}{2}}}}}} }  \approx \frac{1}{{{\epsilon ^2}}}\int {\int_{\mathcal A} {\frac{1}{{{{\left( {1 - 2 y \Phi - 2 z \Omega + {y^2} + {z^2}} \right)}^{\frac{3}{2}}}}}dydz} }\notag \\
&\mathop  = \limits^{\left( a \right)} \frac{1}{{{\epsilon ^2}}}\int_{ - \frac{{{M_y}\epsilon }}{2}}^{\frac{{{M_y}\epsilon }}{2}} {\frac{{\frac{{\frac{{{M_z}\epsilon }}{2} - \cos \theta }}{{\sqrt {{{\left( {y - \sin \theta \sin \phi } \right)}^2} + {{\sin }^2}\theta {{\cos }^2}\phi  + {{\left( {\frac{{{M_z}\epsilon }}{2} - \cos \theta } \right)}^2}} }}}}{{{{\left( {y - \sin \theta \sin \phi } \right)}^2} + {{\sin }^2}\theta {{\cos }^2}\phi }} + \frac{{\frac{{\frac{{{M_z}\epsilon }}{2} + \cos \theta }}{{\sqrt {{{\left( {y - \sin \theta \sin \phi } \right)}^2} + {{\sin }^2}\theta {{\cos }^2}\phi  + {{\left( {\frac{{{M_z}\epsilon }}{2} + \cos \theta } \right)}^2}} }}}}{{{{\left( {y - \sin \theta \sin \phi } \right)}^2} + {{\sin }^2}\theta {{\cos }^2}\phi }}dy} \notag \\
&\mathop  = \limits^{\left( b \right)} \frac{1}{{{\epsilon ^2}\sin \theta \cos \phi }}\left[ {\arctan \left( {\frac{{\left( {\frac{{{M_y}\epsilon }}{2} - \sin \theta \sin \phi } \right)\left( {\frac{{{M_z}\epsilon }}{2} - \cos \theta } \right)}}{{\sin \theta \cos \phi \sqrt {{{\sin }^2}\theta {{\cos }^2}\phi  + {{\left( {\frac{{{M_y}\epsilon }}{2} - \sin \theta \sin \phi } \right)}^2} + {{\left( {\frac{{{M_z}\epsilon }}{2} - \cos \theta } \right)}^2}} }}} \right)} \right. +  \notag\\
&\ \ \ \ \ \ \ \ \ \ \ \ \ \ \ \ \ \ \ \ \ \arctan \left( {\frac{{\left( {\frac{{{M_y}\epsilon }}{2} + \sin \theta \sin \phi } \right)\left( {\frac{{{M_z}\epsilon }}{2} - \cos \theta } \right)}}{{\sin \theta \cos \phi \sqrt {{{\sin }^2}\theta {{\cos }^2}\phi  + {{\left( {\frac{{{M_y}\epsilon }}{2} + \sin \theta \sin \phi } \right)}^2} + {{\left( {\frac{{{M_z}\epsilon }}{2} - \cos \theta } \right)}^2} } }}} \right) +  \notag\\
&\ \ \ \ \ \ \ \ \ \ \ \ \ \ \ \ \ \ \ \ \ \arctan \left( {\frac{{\left( {\frac{{{M_y}\epsilon }}{2} - \sin \theta \sin \phi } \right)\left( {\frac{{{M_z}\epsilon }}{2} + \cos \theta } \right)}}{{\sin \theta \cos \phi \sqrt {{{\sin }^2}\theta {{\cos }^2}\phi  + {{\left( {\frac{{{M_y}\epsilon }}{2} - \sin \theta \sin \phi } \right)}^2} + {{\left( {\frac{{{M_z}\epsilon }}{2} + \cos \theta } \right)}^2} } }}} \right) +  \notag\\
&\ \ \ \ \ \ \ \ \ \ \ \ \ \ \ \ \ \ \ \ \ \left. {\arctan \left( {\frac{{\left( {\frac{{{M_y}\epsilon }}{2} + \sin \theta \sin \phi } \right)\left( {\frac{{{M_z}\epsilon }}{2} + \cos \theta } \right)}}{{\sin \theta \cos \phi \sqrt {{{\sin }^2}\theta {{\cos }^2}\phi  + {{\left( {\frac{{{M_y}\epsilon }}{2} + \sin \theta \sin \phi } \right)}^2} + {{\left( {\frac{{{M_z}\epsilon }}{2} + \cos \theta } \right)}^2} } }}} \right)} \right].
\end{align}
\hrulefill % %这里有一条线，如果你想要
\vspace*{4pt} %留空白，可自己调整
\end{figure*}
Furthermore, by substituting \eqref{Integrate1} into \eqref{singleUserSNRSummation}, the resulting SNR in \eqref{singleUserapproxiamtionSNR} can be obtained. The proof of Theorem~\ref{singleUserSNRapproximationTheorem} is thus completed.

\section{Proof of Lemma~\ref{alternativeExpressionSNRtheta=pi/2lemma}}\label{proofOfalternativeExpressionSNRtheta=pi/2lemma}
Depending on whether the projection $O'$ of the user location on the $y$-axis lies within $\left[ { - \frac{{{L_y}}}{2},\frac{{{L_y}}}{2}} \right]$ or not, the proof of Lemma~\ref{alternativeExpressionSNRtheta=pi/2lemma} involves the following three cases.

\emph{Case 1}:  $ - \frac{{{L_y}}}{2} \le r\sin \phi  \le \frac{{{L_y}}}{2}$, i.e., the projection $O'$ is located within the line segment $D_2D_3$, as illustrated in Fig.~\ref{angleIllustrationSpecialCaseThetapi2}(a). In this case, it is observed from Fig.~\ref{angleIllustrationSpecialCaseThetapi2}(b) that $\tan {\eta _1} = \frac{{{L_y}/2 - r\sin \phi }}{{r\cos \phi }}$ and $\tan {\eta _2} = \frac{{{L_y}/2 + r\sin \phi }}{{r\cos \phi }}$. Besides, as can be seen from Fig.~\ref{angleIllustrationSpecialCaseThetapi2}(a), the line segment $D_1D_2$ is perpendicular to $OD2$. Thus, $\sin {\beta _1} = \frac{{\left| {{D_1}{D_2}} \right|}}{{\sqrt {{{\left| {{D_1}{D_2}} \right|}^2} + {{\left| {O{D_2}} \right|}^2}} }} = \frac{{{L_z}/2}}{{\sqrt {{r^2}{{\cos }^2}\phi  + {{\left( {{L_y}/2 - r\sin \phi } \right)}^2} + {{\left( {{L_z}/2} \right)}^2}} }}$. Similarly, $\sin {\beta _2} =\frac{{\left| {{D_3}{D_4}} \right|}}{{\sqrt {{{\left| {{D_3}{D_4}} \right|}^2} + {{\left| {O{D_3}} \right|}^2}} }} = \frac{{{L_z}/2}}{{\sqrt {{r^2}{{\cos }^2}\phi  + {{\left( {{L_y}/2 + r\sin \phi } \right)}^2} + {{\left( {{L_z}/2} \right)}^2}} }}$. As a result, \eqref{singleUserapproxiamtionSNRtheta=pi/2} is equivalently expressed as
\begin{equation}\label{alternativeExpressionSNRtheta=pi/2Case1}
\gamma  = \frac{{\xi \bar P}}{{2\pi }}\left[ {\arctan \left( {\tan {\eta _1}\sin {\beta _1}} \right)} \right.\left. { + \arctan \left( {\tan {\eta _2}\sin {\beta _2}} \right)} \right].
\end{equation}

\emph{Case 2}: $r\sin \phi  > \frac{{{L_y}}}{2}$, i.e., the projection $O'$ is above $D_2$. It is not difficult to see that only $\tan {\eta _1}$ changes compared to Case 1, which is given by $\tan {\eta _1} = \frac{{r\sin \phi  - {L_y}/2}}{{r\cos \phi }}$. Thus, by changing the sign of $\tan {\eta _1}$ in \eqref{alternativeExpressionSNRtheta=pi/2Case1}, an equivalent expression of \eqref{singleUserapproxiamtionSNRtheta=pi/2} can be obtained.

\emph{Case 3}: $r\sin \phi  <  - \frac{{{L_y}}}{2}$, i.e., the projection $O'$ is below $D_3$. In this case, the change of $\tan {\eta _2}$ occurs compared to Case 1, which is given by $\tan {\eta _2} =  - \frac{{{L_y}/2 + r\sin \phi }}{{r\cos \phi }}$. Similarly, the alternative expression of \eqref{singleUserapproxiamtionSNRtheta=pi/2} is obtained by changing the sign of $\tan {\eta _2}$ in \eqref{alternativeExpressionSNRtheta=pi/2Case1}.

By summarizing the above three cases, Lemma~\ref{alternativeExpressionSNRtheta=pi/2lemma} is proved.

\section{Proof of Lemma~\ref{ReduceToUPW}}\label{proofOfReduceToUPWLemma}
When $r\Psi  \gg {L_y}$ and $r\Psi  \gg {L_z}$, we have $\frac{{{L_y}}}{{2r\Psi }} \ll 1$ and $\frac{{{L_z}}}{{2r\Psi }} \ll 1$. By using the fact that $\arctan x \approx x$ for $\left| x \right| \ll 1$ and the condition $\frac{{\frac{\Phi }{\Psi }\frac{\Omega }{\Psi }}}{{\sqrt {1 + {{\left( {\frac{{{L_y}}}{{2r\Psi }} \pm \frac{\Phi }{\Psi }} \right)}^2} + {{\left( {\frac{{{L_z}}}{{2r\Psi }} \pm \frac{\Omega }{\Psi }} \right)}^2}} }} \ll 1$, \eqref{singleUserapproxiamtionSNR} can be approximated as
\begin{equation}\label{singleUserapproxiamtionSNRFarField1}
{\small
\begin{aligned}
&\gamma  \approx \\
&\frac{{\xi \bar P}}{{4\pi }}\left[ {{U_1}\left( {\frac{{{L_y}}}{{2r\Psi }} - \frac{{\Phi }}{{\Psi }},\frac{{{L_z}}}{{2r\Psi }} - \frac{{\Omega }}{{\Psi }}} \right)} \right. + {U_1}\left( {\frac{{{L_y}}}{{2r\Psi }} + \frac{{\Phi }}{{\Psi }},\frac{{{L_z}}}{{2r\Psi }} - \frac{{\Omega }}{{\Psi }}} \right)\\
&\left. { + {U_1}\left( {\frac{{{L_y}}}{{2r\Psi }} - \frac{{\Phi }}{{\Psi }},\frac{{{L_z}}}{{2r\Psi }} + \frac{{\Omega }}{{\Psi }}} \right) + {U_1}\left( {\frac{{{L_y}}}{{2r\Psi }} + \frac{{\Phi }}{{\Psi }},\frac{{{L_z}}}{{2r\Psi }} + \frac{{\Omega }}{{\Psi }}} \right)} \right],
\end{aligned}}
\end{equation}
where ${U_1}\left( {x,y} \right) = \frac{{xy}}{{\sqrt {1 + {x^2} + {y^2}} }}$. Due to the similar form of the four terms inside the bracket, we first express the denominator of the first term as a function of $u \buildrel \Delta \over = \frac{{{L_y}}}{{2r\Psi }}$ and $v \buildrel \Delta \over = \frac{{{L_z}}}{{2r\Psi }}$, defined as $g\left( {u,v} \right) \buildrel \Delta \over = \sqrt {1 + {{\left( {u - \frac{{\Phi }}{{\Psi }}} \right)}^2} + {{\left( {v - \frac{{\Omega }}{{\Psi }}} \right)}^2}} $. By applying the first-order Taylor approximation to $g\left( {u,v} \right)$ for small $u$ and $v$, it follows that
\begin{equation}\label{gyzfirst-orderTaylorApproximation}
g\left( {u,v} \right) \approx B - \frac{{\frac{{\Phi }}{{\Psi }}u}}{B} - \frac{{\frac{{\Omega }}{{\Psi }}v}}{B},\ B \buildrel \Delta \over =  \sqrt {1 + \frac{{{{\Phi }^2}}}{{{{\Psi }^2}}} + \frac{{{{\Omega }^2}}}{{{{\Psi }^2}}}}.
\end{equation}
By following the similar procedure, the first-order Taylor approximation to the denominator of other three terms in \eqref{singleUserapproxiamtionSNRFarField1} can be obtained. Thus, \eqref{singleUserapproxiamtionSNRFarField1} can be approximated as
\begin{equation}\label{singleUserapproxiamtionSNRFarField2}
\begin{aligned}
\gamma  \approx \frac{{\xi \bar P}}{{4\pi }}\left[ {{U_2}\left( { - \frac{{\Phi }}{{\Psi }}, - \frac{{\Omega }}{{\Psi }}} \right)} \right. + {U_2}\left( {\frac{{\Phi }}{{\Psi }}, - \frac{{\Omega }}{{\Psi }}} \right)\\
\left. { + {U_2}\left( { - \frac{{\Phi }}{{\Psi }},\frac{{\Omega }}{{\Psi }}} \right) + {U_2}\left( {\frac{{\Phi }}{{\Psi }},\frac{{\Omega }}{{\Psi }}} \right)} \right],
\end{aligned}
\end{equation}
where ${U_2}\left( {x,y} \right){\rm{ }} \buildrel \Delta \over = \frac{{\left( {\frac{{{L_y}}}{{2r\Psi }} + x} \right)\left( {\frac{{{L_z}}}{{2r\Psi }} + y} \right)}}{{B + \frac{x}{B}\frac{{{L_y}}}{{2r\Psi }} + \frac{y}{B}\frac{{{L_z}}}{{2r\Psi }}}}$. Furthermore, the sum of ${{U_2}\left( { - \frac{{\Phi }}{{\Psi }}, - \frac{{\Omega }}{{\Psi }}} \right)}$ and ${{U_2}\left( {\frac{{\Phi }}{{\Psi }},\frac{{\Omega }}{{\Psi }}} \right)}$ is given by
\begin{equation}
 \hspace{-0.1cm}
{\small
\begin{aligned}
&{U_2}\left( { - \frac{{\Phi }}{{\Psi }}, - \frac{{\Omega }}{{\Psi }}} \right) + {U_2}\left( {\frac{{\Phi }}{{\Psi }},\frac{{\Omega }}{{\Psi }}} \right)\\
& = \frac{{\left( {\frac{{{L_y}}}{{2r\Psi }} - \frac{{\Phi }}{{\Psi }}} \right)\left( {\frac{{{L_z}}}{{2r\Psi }} - \frac{{\Omega }}{{\Psi }}} \right)}}{{B - \frac{{\Phi }}{{B\Psi }}\frac{{{L_y}}}{{2r\Psi }} - \frac{{\Omega }}{{B\Psi }}\frac{{{L_z}}}{{2r\Psi }}}} + \frac{{\left( {\frac{{{L_y}}}{{2r\Psi }} + \frac{{\Phi }}{{\Psi }}} \right)\left( {\frac{{{L_z}}}{{2r\Psi }} + \frac{{\Omega }}{{\Psi }}} \right)}}{{B + \frac{{\Phi }}{{B\Psi }}\frac{{{L_y}}}{{2r\Psi }} + \frac{{\Omega }}{{B\Psi }}\frac{{{L_z}}}{{2r\Psi }}}}\\
& = 2\frac{{\left( {B - \frac{{{{\Omega }^2}}}{{B{{\Psi }^2}}} - \frac{{{{\Phi }^2}}}{{B{{\Psi }^2}}}} \right)\frac{{{L_y}{L_z}}}{{4{r^2}{{\Psi }^2}}} + B\frac{{\Phi \Omega }}{{{{\Psi }^2}}} - \frac{{\Phi \Omega }}{{B{{\Psi }^2}}}\left( {\frac{{L_y^2}}{{4{r^2}{{\Psi }^2}}} + \frac{{L_z^2}}{{4{r^2}{{\Psi }^2}}}} \right)}}{{{B^2} - {{\left( {\frac{{\Phi }}{{B\Psi }}\frac{{{L_y}}}{{2r\Psi }} + \frac{{\Omega }}{{B\Psi }}\frac{{{L_z}}}{{2r\Psi }}} \right)}^2}}}\\
&\mathop  \approx \limits^{\left( a \right)} 2\frac{{\left( {B - \frac{{{{\Omega }^2}}}{{B{{\Psi }^2}}} - \frac{{{{\Phi }^2}}}{{B{{\Psi }^2}}}} \right)\frac{{{L_y}{L_z}}}{{4{r^2}{{\Psi }^2}}} + B\frac{{\Phi \Omega }}{{{{\Psi }^2}}} - \frac{{\Phi \Omega }}{{B{{\Psi }^2}}}\left( {\frac{{L_y^2}}{{4{r^2}{{\Psi }^2}}} + \frac{{L_z^2}}{{4{r^2}{{\Psi }^2}}}} \right)}}{{{B^2}}},
\end{aligned}}
\end{equation}
where ${\left( a \right)}$ follows from the fact that $\frac{{{L_y}}}{{2r\Psi }} \ll 1$ and $\frac{{{L_z}}}{{2r\Psi }} \ll 1$. Furthermore, the sum of ${U_2}\left( {\frac{{\Phi }}{{\Psi }}, - \frac{{\Omega }}{{\Psi }}} \right)$ and ${{U_2}\left( { - \frac{{\Phi }}{{\Psi }},\frac{{\Omega }}{{\Psi }}} \right)}$ can be similarly obtained. By combining the four terms, we obtain
\begin{equation}
\begin{aligned}
\gamma  &\approx \frac{{\xi \bar P}}{{4\pi }}4\frac{{\left( {B - \frac{{{{\Omega }^2}}}{{B{{\Psi }^2}}} - \frac{{{{\Phi }^2}}}{{B{{\Psi }^2}}}} \right)\frac{{{L_y}{L_z}}}{{4{r^2}{{\Psi }^2}}}}}{{{B^2}}}\\
 &= \frac{{\bar PMA}}{{4\pi {r^2}{{\Psi }^2}}}\frac{{\left( {{B^2} - \frac{{{{\Omega }^2}}}{{{{\Psi }^2}}} - \frac{{{{\Phi }^2}}}{{{{\Psi }^2}}}} \right)}}{{{B^3}}}\\
 &= \frac{{\bar PMA}}{{4\pi {r^2}{{\Psi }^2}}}\frac{1}{{{{\left( {1 + \frac{{{{\Omega }^2}}}{{{{\Psi }^2}}} + \frac{{{{\Phi }^2}}}{{{{\Psi }^2}}}} \right)}^{\frac{3}{2}}}}} = \frac{{\bar PMA}}{{4\pi {r^2}{{\Psi }^2}}}{{\Psi }^3} = \frac{{\bar PMA}}{{4\pi {r^2}}}\Psi.
\end{aligned}
\end{equation}
 This thus completes the proof of Lemma~\ref{ReduceToUPW}.

\section{Proof of Lemma~\ref{TransformUniformLinearArray}}\label{proofOfTransformUniformLinearArray}
By substituting ${M_y} = 1$ and $M = {M_z}$ into \eqref{singleUserapproxiamtionSNR}, we have
\begin{equation}\label{ULAExpression}
\begin{aligned}
\gamma  = \frac{{\xi \bar P}}{{4\pi }}\left[ {U\left( {\frac{d}{{2r}} - \Phi ,\frac{{{L_z}}}{{2r}} - \Omega } \right)} \right. + U\left( {\frac{d}{{2r}} + \Phi ,\frac{{{L_z}}}{{2r}} - \Omega } \right)\\
\left. { + U\left( {\frac{d}{{2r}} - \Phi ,\frac{{{L_z}}}{{2r}} + \Omega } \right) + U\left( {\frac{d}{{2r}} + \Phi ,\frac{{{L_z}}}{{2r}} + \Omega } \right)} \right].
\end{aligned}
\end{equation}
We first express the first term ${U\left( {\frac{d}{{2r}} - \Phi ,\frac{{{L_z}}}{{2r}} - \Omega } \right)}$ inside the bracket as a function of $\epsilon$, defined as $h\left( \epsilon  \right) \buildrel \Delta \over = \arctan \left( {\frac{{\left( {\epsilon /2 - \Phi } \right)E}}{{\sqrt {F + {\epsilon ^2}/4 - \Phi \epsilon } }}} \right)$, where $E \buildrel \Delta \over = \frac{{{L_z}}}{{2r\Psi }} - \frac{\Omega }{\Psi }$ and $F \buildrel \Delta \over = 1 + \frac{{L_z^2}}{{4{r^2}}} - \frac{{{L_z}}}{r}\Omega $. By applying the first-order Taylor approximation to $h\left( \epsilon \right)$ for small $\epsilon$, it follows that
\begin{equation}\label{ULAExpressionfirst-orderTaylorapproximation}
\begin{aligned}
h\left( \epsilon  \right) &\approx h\left( 0 \right) + h'\left( 0 \right)\epsilon \\
 &=  - \arctan \left( {\frac{{\Phi E}}{{\sqrt F }}} \right) + \frac{E}{{\sqrt F }}\frac{{\left( {F - {\Phi ^2}} \right)}}{{\left( {F + {E^2}{\Phi ^2}} \right)}}\frac{\epsilon }{2}\\
 &=  - \arctan \left( {\frac{{\Phi E}}{{\sqrt F }}} \right) + \frac{E}{{\sqrt F }}\frac{{F - {\Phi ^2}}}{{F + \left( {F - {{\sin }^2}\theta } \right){{\tan }^2}\phi }}\frac{\epsilon }{2}\\
 &=  - \arctan \left( {\frac{{\Phi E}}{{\sqrt F }}} \right) + \frac{E}{{\sqrt F }}\frac{{F - {\Phi ^2}}}{{\left( {F - {\Phi ^2}} \right){{\sec }^2}\phi }}\frac{\epsilon }{2}\\
 &=  - \arctan \left( {\frac{{\Phi E}}{{\sqrt F }}} \right) + \frac{{E\epsilon }}{{2\sqrt F }}{\cos ^2}\phi.
\end{aligned}
\end{equation}
By following the similar procedure for other three terms inside the bracket of \eqref{ULAExpression}, we have
\begin{equation}\label{ULAExpressionfirst-orderTaylorapproximation2}
\begin{aligned}
&\gamma  \approx \frac{{\xi \bar P}}{{4\pi }}\epsilon \left( {\frac{{\frac{{{L_z}}}{{2r\Psi }} - \frac{{\Omega }}{{\Psi }}}}{{\sqrt {1 - \frac{{{L_z}}}{r}\Omega  + \frac{{L_z^2}}{{4{r^2}}}} }} + \frac{{\frac{{{L_z}}}{{2r\Psi }}{\rm{ + }}\frac{{\Omega }}{{\Psi }}}}{{\sqrt {1 + \frac{{{L_z}}}{r}\Omega  + \frac{{L_z^2}}{{4{r^2}}}} }}} \right){\cos ^2}\phi \\
& = \frac{{\bar PA\cos \phi }}{{4\pi dr\sin \theta }}\left( {\frac{{\frac{{Md}}{2} - r\cos \theta }}{{\sqrt {{{\left( {\frac{{Md}}{2} - r\cos \theta } \right)}^2} + {r^2}{{\sin }^2}\theta } }}} \right.\\
&\ \ \ \ \ \ \ \ \ \ \ \ \ \ \ \ \ \ \ \ \ \ \ \ \ + \left. {\frac{{\frac{{Md}}{2} + r\cos \theta }}{{\sqrt {{{\left( {\frac{{Md}}{2} + r\cos \theta } \right)}^2} + {r^2}{{\sin }^2}\theta } }}} \right).
\end{aligned}
\end{equation}
It is observed from Fig.~\ref{ULAsystemModel} that the two terms into the bracket of \eqref{ULAExpressionfirst-orderTaylorapproximation2} are just the sine of ${\alpha _1}\left( M \right)$ and ${\alpha _2}\left( M \right) $, where ${\alpha _1}\left( M \right) \buildrel \Delta \over = \arctan \left( {\frac{{Md/2 - r\cos \theta }}{{r\sin \theta }}} \right)$ and ${\alpha _{\rm{2}}}\left( M \right) \buildrel \Delta \over = \arctan \left( {\frac{{Md/2 + r\cos \theta }}{{r\sin \theta }}} \right)$, respectively. Thus, it follows that
\begin{equation}
 \gamma  = \frac{{\bar PA\cos \phi }}{{4\pi dr\sin \theta }}\left[ {\sin \left( {{\alpha _1}\left( M \right)} \right) + \sin \left( {{\alpha _{\rm{2}}}\left( M \right)} \right)} \right].
 \end{equation}
The proof of Lemma~\ref{TransformUniformLinearArray} is thus completed.
\end{appendices}
%\ifCLASSOPTIONcaptionsoff
%  \newpage
%\fi

\bibliographystyle{IEEEtran}
\bibliography{refXLMIMO}

\end{document}